# Energy-Efficient and Fast Memristor-based Serial Multipliers Applicable in Image Processing

Seyed Erfan Fatemieh†, Bahareh Bagheralmoosavi†, and Mohammad Reza Reshadinezhad†,*

## Abstract

Memristive Processing In-Memory (PIM) is one of the promising techniques for overcoming the Von-Neumann bottleneck. Reduction of data transfer between processor and memory and data processing by memristors in data-intensive applications reduces energy consumption and processing time. Multipliers are one of the fundamental arithmetic circuits that play a significant role in data-intensive processing applications. The computational complexity of multipliers has turned them into one of the arithmetic circuits affecting PIM's efficiency and energy consumption, for example, in convolution operations. Serial material implication (IMPLY) logic design is one of the methods of implementing arithmetic circuits by applying emerging memristive technology that enables PIM in the structure of crossbar arrays. The authors propose unsigned and signed array multipliers using serial IMPLY logic in this paper. The proposed multipliers have improved significantly compared to State-Of-the Art (SOA) by applying the proposed Partial Product Units (PPUs) and overlapping computational steps. The number of computational steps, energy consumption, and required memristors of the proposed 8-bit unsigned array multiplier are improved by up to 36%, 31%, and 47% compared to the classic designs. The proposed 8-bit signed multiplier has also improved the computational steps, energy consumption, and required memristors by up to 59%, 54%, and 45%. The performance of the proposed multipliers in the applications of Gaussian blur and edge detection is also investigated, and the simulation results have shown an improvement of 31% in energy consumption and 33% in the number of computational steps in these applications.

## 1 Introduction

In the era of expanding data-intensive applications, industry and researchers face challenges in designing efficient processing systems. Image processing, the Internet of Things (IoT), and machine learning are among the most critical data-intensive applications [1]. For a long time, the Von-Neumann architecture has been the conventional architecture for these applications, which separates the processing units and the memory (data storage) at a distance. Data movement between processing units and memory in today's processing systems is associated with several challenges, such as reducing efficiency and increasing energy consumption, known as "Von-Neumann's bottleneck" [1-6]. Since the 80s, the performance of processors has doubled every two years, while the performance of memory has doubled every 10 years [7, 8]. However, due to the limitations of the power wall and transistors' scalability, processor performance growth has slowed in recent years, with performance doubling almost every 20 years [8-10]. Increased leakage currents lead to higher power consumption and reduced efficiency growth [11-13]. Also, the energy consumption of data movement between memory and the processing unit is challenging [4]. Various solutions have been proposed to improve energy consumption and overcome the Von-Neumann bottleneck, including near-memory processing (placing several cache levels close to the processor or 3D stacking of Dynamic Random Access Memory (DRAM) on the processor) and PIM [1, 2, 5, 6]. In the PIM paradigm, the memory array can process data and perform logical and arithmetic operations besides storing data. Conventional memories, including Static RAM (SRAM) and emerging nonlinear nanotechnologies such as memristors (Resistive RAM (RRAM)), are being applied in academia and industry for integration with arithmetic and logic units [2, 14].

† S. E. Fatemieh (erfanfatemieh@eng.ui.ac.ir), B. Bagheralmoosavi (b.mousavi@eng.ui.ac.ir), and M. R. Reshadinezhad (m.reshadinezhad@eng.ui.ac.ir) are with the Department of Computer Architecture, Faculty of Computer Engineering, University of Isfahan, Isfahan 8174673441, Iran.

* Corresponding author: Mohammad Reza Reshadinezhad.

RRAM is known for its high retention time in different conditions, nanometric dimensions, compatibility with the Complementary Metal Oxide Semiconductor (CMOS) manufacturing process, high switching speed, and low energy consumption [1-5, 15]. Memristors have been commercialized by Intel, Fujitsu, Panasonic, etc., and researchers are looking to replace them with DRAM and flash memory cells [3, 15]. The resistance of the RRAM cell changes to Low Resistance State (LRS) and High Resistance State (HRS) when voltage/current is applied in the direction of polarity or vice versa, and otherwise, its resistance value remains constant [1-4]. Several methods, including IMPLY, Fast and Energy-Efficient Logic (FELIX), and Memristor Aided LoGIC (MAGIC), have been proposed to implement and execute logical and arithmetic instructions using memristors as emerging memory cells [16-18]. IMPLY is a stateful logic design method that represents logic zero and one by HRS and LRS, respectively [4, 5, 18]. Stateful logic methods can help overcome the Von-Neumann bottleneck when designing arithmetic circuits for PIM. This approach can reduce energy consumption by minimizing data transmission [5, 20]. With its simple and compact structure, IMPLY logic is well-suited for coordination with crossbar arrays.

Several applied logic gates and arithmetic operations in data-intensive applications, including adders and multipliers, use stateful logic methods such as IMPLY [1, 16, 20-25]. The structural complexity of multipliers is more than that of adder circuits. Computing the product of two numbers requires more time (number of computational cycles) and energy consumption than computing their sum. Considering the importance of multipliers as fundamental arithmetic circuits in applications such as convolution and multiply-accumulate (MAC) operations, reducing the hardware complexity and computational delay directly affects overall efficiency in processing these applications [11, 26]. Several multipliers have been introduced using the IMPLY logic design method based on parallel architecture to reduce the computational steps [24, 25]. However, this topology increases hardware complexity and there is no proper coordination between parallel structure and crossbar arrays [1, 2].

Signed and unsigned multiplications are among the main arithmetic operations in data processing applications. Designing stateful memristive multipliers compatible with the structure of crossbar arrays can be a priority for PIM. Several algorithms and methods have been introduced to implement energy/area efficient fast unsigned and signed multipliers regardless of the implementation technology. Dadda, array, and add & shift multipliers are commonly used for unsigned multiplication [27]. Additionally, signed multiplication is often carried out using the Baugh-Wooley multiplier with the Dadda reduction tree, radix-2 Booth multiplier, and array multiplier [27]. In this paper, the authors' main goal is to present IMPLY-based serial unsigned and signed multipliers compatible with crossbar arrays to reduce the computational steps and energy consumption compared to SOA. The main contributions of this paper are:

1. Implementation of the classic add & shift and classic Dadda multipliers by applying the IMPLY logic method in serial topology, calculating the number of memristors and computational steps, and estimating the energy consumption of the n-bit structure,
2. Implementation of add & shift, radix-2 Booth, Baugh-Wooley (with Dadda reduction tree) as classic signed multipliers using IMPLY logic in serial topology and calculating circuit analysis metrics of the n-bit structures,
3. Implementation of classic unsigned and signed array multipliers with Carry Save Adder (CSA) tree based on the IMPLY method in serial architecture and calculating the number of computational steps, the number of memristors, and the energy consumption of the n-bit structure,
4. Proposing nine serial IMPLY-based PPUs to be applied in unsigned and signed array multipliers based on the CSA tree to reduce the number of computational steps and the energy consumption of the n-bit serial multiplier,
5. Calculation of the number of computational steps, memristors, and the energy consumption of the proposed improved n-bit unsigned and signed array multipliers and
6. Investigating the functionality of the proposed multipliers and SOA in Gaussian blur and edge detection applications and evaluating the circuit analysis metrics improvement.

The rest of the article is divided into four sections. The second section describes IMPLY as a stateful logic method, common implementation architectures based on it, and the reasons for selecting serial topology. Then, the basic cells required to implement IMPLY-based classic multipliers are introduced in subsection 2.2. Serial multipliers introduced based on the IMPLY design method in recent years are assessed in the last subsection of Section 2, along with the implementation details of classic unsigned and signed multipliers. Classic unsigned and signed array multipliers, improved PPUs, and proposed array multipliers are reported in the third section. The fourth section presents circuit-level simulation results, their analysis and evaluation, and a comparison of the proposed circuit with previous works. In the last subsection of section 4, the results of application-level simulation (Gaussian blur and edge detection) are reported, along with the evaluation and comparison. Finally, in the fifth section, the article concludes.

# 2 Previous Works

## 2.1 IMPLY logic and its implementation architectures

IMPLY logic operation ($p \rightarrow q \equiv \neg p + q$) can be implemented by two memristors, $p$ and $q$, connected to the ground through a resistor ($R_G$), as shown in Figure 1(a). By simultaneously applying voltages $V_{COND}$ and $V_{SET}$ to memristors $p$ and $q$, respectively, the result of $p \rightarrow q$ is stored in memristor $q$. The binary value stored in each memristor can be changed to the logic zero value (HRS) by applying the $V_{RESET}$ voltage to the memristor. The truth table of $p \rightarrow q$ is written in Table 1. Two conditions must be met for the IMPLY gate, based on two memristors $p$ and $q$, to function properly: 1) $V_{COND} < V_{SET}$, and 2) $R_{ON} \ll R_G \ll R_{OFF}$ [5]. Various logic gates, adders, and multipliers are implemented based on IMPLY logic [4, 5, 16, 20-23]. Four different architectures can be applied to implement different IMPLY-based circuits. In the serial architecture, the memristors (input, output, and work) are all placed in a row or column, and only one IMPLY operation is executed in each computational cycle [21]. The structure of serial architecture is shown in Figure 1(b). In parallel architecture, the input, output, and work memristors are placed in different rows/columns and are connected to the $R_G$ resistor of each row/column by CMOS switches [21]. Also, rows/columns can be connected through CMOS switches. Assume no dependency between memristors in different rows/columns in each cycle. In that case, one IMPLY operation is executed in each row/column based on the step order of the implementation algorithm [21, 23]. Semi-serial and semi-parallel architectures have also been proposed to trade-off processing time and hardware complexity between serial and parallel architectures [21, 22].

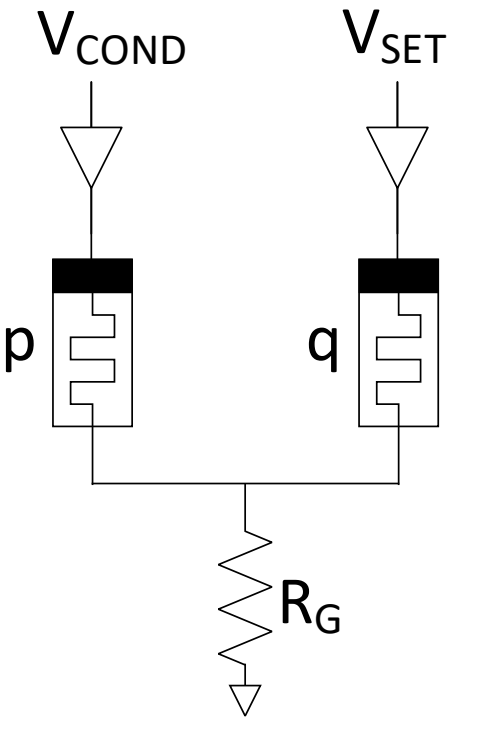

(a)

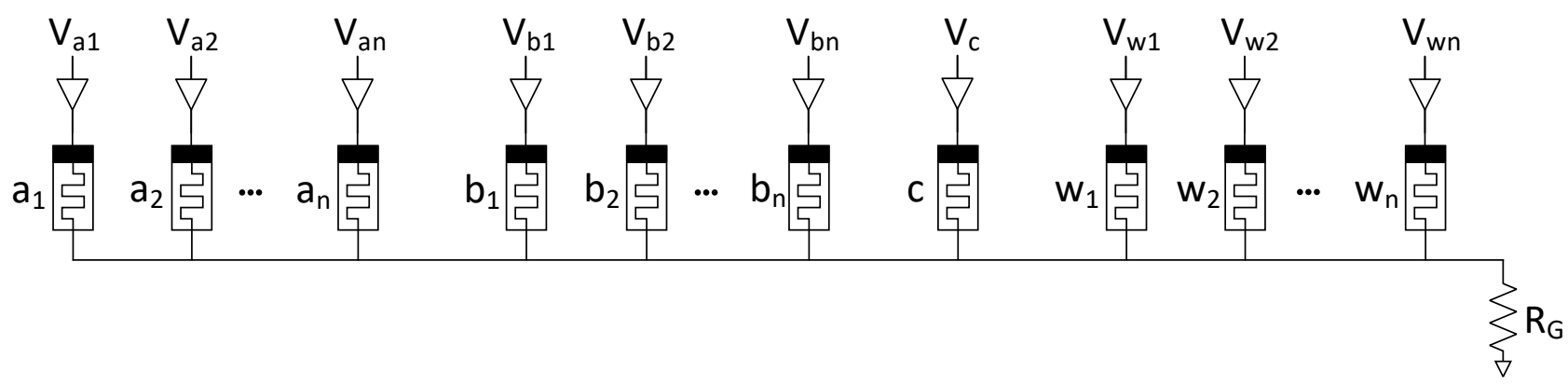

(b)

Figure 1: (a) Memristive IMPLY gate, and (b) Memristive IMPLY-based serial architecture [28].

Table 1: The IMPLY logic gate's truth table.

| **Case** | $p$ | $q$ | $p \rightarrow q$ |
|---|---|---|---|
| 1 | 0 | 0 | 1 |
| 2 | 0 | 1 | 1 |
| 3 | 1 | 0 | 0 |
| 4 | 1 | 1 | 1 |

The coordination of the applied architecture and the design method with the crossbar array structure is one of the essential features to consider when designing memristive arithmetic circuits for PIM. Implementing arithmetic circuits based on parallel architecture is not recommended due to the complexity and incompatibility with the crossbar array [2, 16]. Semi-serial and semi-parallel architectures have moderate compatibility with the memristive crossbar array [2]. The proposed circuits of this article and the reviewed SOA in the following subsections are only based on the serial architecture due to its simplicity and compatibility with the crossbar array, and other architectures will not be reviewed.

## 2.2 IMPLY-based components of the multiplier

After a thorough investigation of standard multipliers, it is determined that three primary stages must be implemented to compute the final product. The first step is Partial Product Generation (PPG). AND and NAND gates can be applied to generate the partial products in unsigned and signed multipliers. In the second stage of implementing the multiplier, Partial Product Reduction (PPR), half adder, full adder, and compressor cells are applied to simplify the multiplication tree. Several algorithms have been introduced to reduce the multiplication tree, one of the most important of which is the Dadda reduction algorithm [27]. In this article, half adder and full adder cells proposed in [23, 28] are applied to reduce the multiplication tree. In the third stage, the final product is calculated by the Ripple Carry Adder (RCA), in which inputs are the output of the second stage. In the structure of add & shift and radix-2 Booth multipliers, multiplexers are needed in addition to full adder cells.

Table 2 summarizes the features of basic circuits required to implement unsigned and signed classic multipliers. The total number of memristors (input/output and work), computational steps, and the possibility of reusing the input memristors to store the output(s) are written for each memristive circuit.

Table 2: Characteristics of basic circuits required to implement classic unsigned and signed multipliers.

| **Circuit** | **No. of Steps** | **Total No. of Memristors** | **Input Memristor Reusability** |
|---|---|---|---|
| NOT($a$) [23] | 2 | 2 | – |
| AND($a$, $b$) [23] | 5 | 3–4 | + |
| NAND($a$, $b$) [23] | 3 | 3 | + |
| Half Adder($a$, $b$) [28] | 12 | 4 | + |
| Full Adder($a$, $b$, $C_{in}$) [23] | 22 | 5 | + |
| Copy($a$) | 4 | 3–4 | + |
| XOR($a$, $b$) [23] | 9 | 4 | + |
| First MUX2:1($a$, $b$, *Select*) | 9 | 6 | – |
| Second MUX2:1($a$, $b$, *Select*) | 7 | 5 | – |

## 2.3 IMPLY-based serial multipliers

In [28], modified IMPLY-based serial 4-bit and 8-bit multipliers are introduced. The authors' main idea in [28] was to present a fast and energy-efficient memristive 4:2 compressor. The outputs of this compressor are computed in only 44 computational steps by applying seven memristors. The 4:2 compressor proposed in [29] needs 52 computational steps and seven memristors, the same as the circuit proposed in [28], to compute the outputs. The PPR and RCA stages of the proposed multipliers in [28] are merged by applying the serial architecture. The n-bit implementation of the multiplier requires $n^2 + 2$ memristors to compute the final product in $27n^2 - 32n$ computational steps [28].
To the best of our knowledge, the only fully serial multiplier based on IMPLY logic has been introduced in [28]. Considering this limitation, the authors have implemented signed and unsigned conventional multipliers by applying the basic cells introduced in subsection 2.2 based on IMPLY logic and serial architecture. So, the proposed unsigned and signed multipliers can be compared with the classic designs. It is tried to use the minimum number of work memristors in addition to the input/output memristors to implement the classic multipliers. Moreover, the authors tried to minimize the steps of the multipliers' implementation algorithms and consider possible overlaps between the computational steps of different applied basic cells, such as the full adder proposed in [23]. The specifications of the presented unsigned and signed multipliers are reported in Tables 3 and 4, respectively. The number of computational steps and required memristors (input/output and work memristors) for implementing the n-bit multipliers are also written in Tables 3 and 4. It should be mentioned that circuits designed in serial architecture do not need CMOS switches for proper functionality. Hence, the number of CMOS switches required by the multipliers and their components is zero.

Table 3: Characteristics of classic unsigned multipliers.

| **Classic Multiplier** | **No. of Steps** | **Total No. of Memristors** |
|---|---|---|
| [28] | $27n^2 - 32n$ | $n^2 + 2$ |
| Dadda | $27n^2 - 32n$ | $n^2 + 2$ |
| Add & Shift | $31n^2 + n + 4$ | $3n + 5$ |

Table 4: Characteristics of classic signed multipliers.

| **Classic Multiplier** | **No. of Steps** | **Total No. of Memristors** |
|---|---|---|
| Radix-2 Booth | $49n^2 + 15n - 4$ | $4n + 8$ |
| Baugh-Wooley (Dadda) | $27n^2 - 24n + 24$ | $n^2 + 2$ |
| Add & Shift | $31n^2 + n + 4$ | $3n + 5$ |

It should be noted that some of the combinational logic circuits applied in classic multipliers are implemented without any optimization and overlapping computational steps. The 2:1 multiplexers and carry generation circuit applied in add & shift and radix-2 Booth multipliers are among these designed combinational circuits.

# 3 Proposed IMPLY-based serial unsigned and signed multipliers

## 3.1 Unsigned serial array multiplier

The array multiplier is known for its well-organized structure. It consists of partial product units (PPUs) made up of full adders and half adders arranged in an array (rows and columns) next to each other. Figure 2 depicts the structure of a 4-bit unsigned array multiplier based on CSA.
The unsigned array multiplier's outputs are obtained by applying four arithmetic blocks. The output's least significant bit (LSB) is calculated using the AND function of the inputs' LSB, $x_0$ and $y_0$. The other output bits of this multiplier are computed by applying three different PPUs ($PPU_1$, $PPU_2$, and $PPU_3$) in addition to a half adder and some full adders. Each of these PPUs' structure are specified with different color in Figure 2. Principally, PPUs are placed in $n$ rows and $n - 1$ diagonal lines in the n-bit unsigned array multiplier. The $PPU_1$ is an arithmetic unit that uses two AND gates to generate partial products and a half adder to add these partial products. The $PPU_2$ is another arithmetic unit consisting of one full adder cell applied to sum up a partial product generated by an AND gate with the outputs of the previous stage's PPUs. Similarly, two AND gates and a full adder cell are the building blocks of the $PPU_3$. 4, 3, and 4 memristors (input and work memristors) are needed to implement the $PPU_1$, $PPU_2$, and $PPU_3$ by applying the serial IMPLY logic, respectively, based on the basic gates and arithmetic cells listed in Table 2.

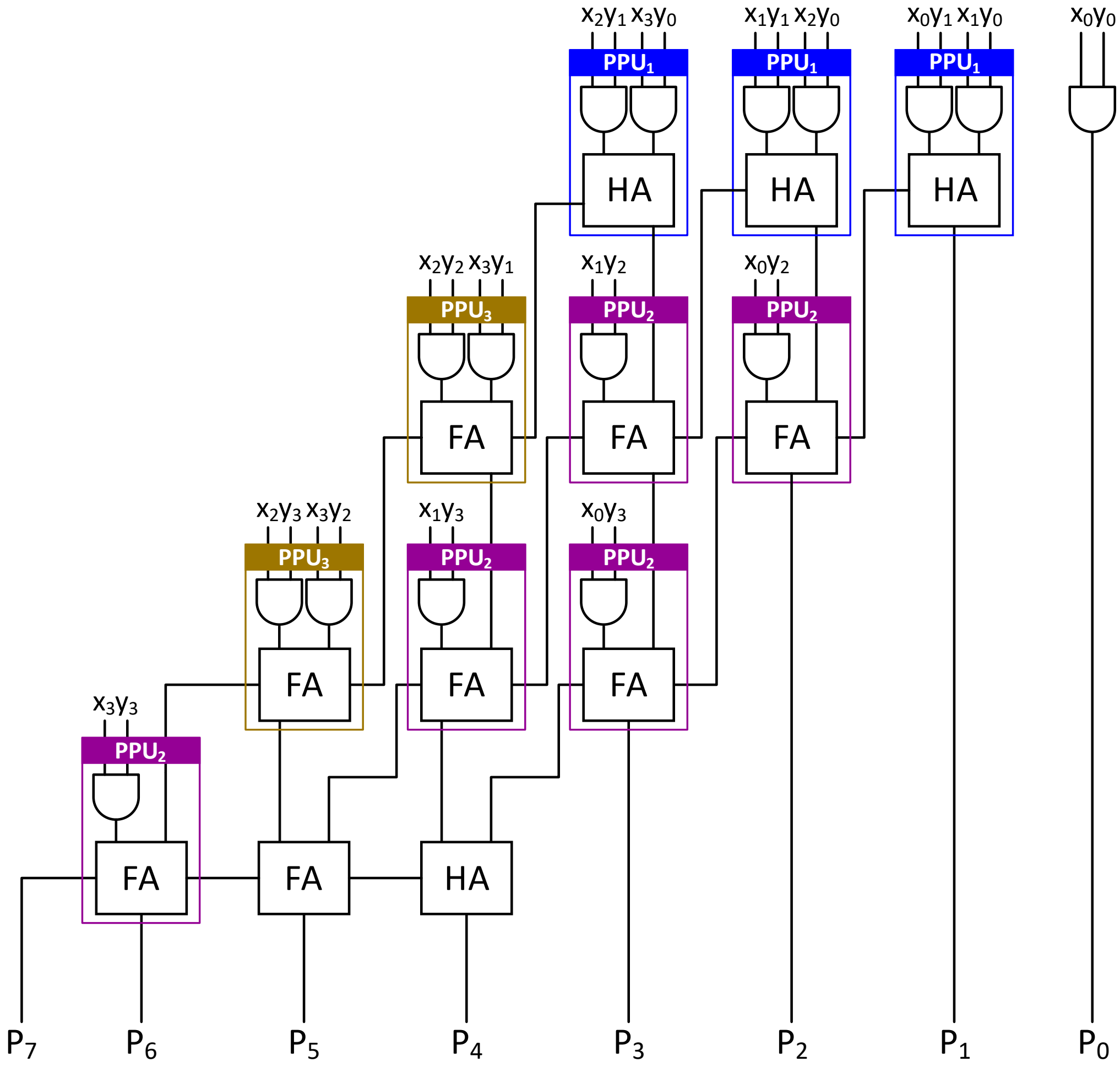

Figure 2: 4-bit unsigned CSA-based array multiplier.

The number of required building blocks for implementing the classic n-bit unsigned array multiplier as well as their computational steps is reported in Table 5. Implementing each building block using IMPLY logic in the serial architecture requires 5 (AND gate) to 32 ($PPU_3$) computational steps, according to the steps reported in Table 5. Therefore, $27n^2 - 32n$ computational steps and $5n - 4$ memristors ($2n$ input memristors and $3n - 4$ work memristors) are required to implement the classic n-bit unsigned array multiplier. The outputs of the classic serial unsigned array multiplier implemented based on IMPLY logic are written in the input memristors, and no extra memristor is needed to store the outputs.

Table 5: Characteristics of classic unsigned array multiplier.

| **Combinational Block** | **No. of Applied Blocks** | **No. of Computational Steps** |
|---|---|---|
| $PPU_1$ | $n - 1$ | 22 |
| $PPU_2$ | $n^2 - 4n + 5$ | 27 |
| $PPU_3$ | $n - 2$ | 32 |
| Half Adder [28] | $n - 1$ | 1 |
| Full Adder [23] | $n - 3$ | 22 |
| AND Gate | 1 | 5 |

Overlapping repeated serial steps in the arithmetic cell's implementation consisting of several primary blocks reduces the computational steps and hardware complexity (reducing the number of memristors) in IMPLY-based serial designs [28]. In $PPU_1$, the outputs of two AND gates are the inputs of the half adder cell. In the classic structure, the number of computational steps of $PPU_1$ equals the sum of the computational steps of two logical AND gates (10 steps) and a half adder cell (12 steps). Thus, the total computational steps of classic $PPU_1$ are 22 steps.

Table 6: Serial IMPLY-based implementation of the AND gate.

| Step | Operation | Equivalent Logic |
|---|---|---|
| **1** | $S_1 = 0$ | FALSE ($S_1$) |
| **2** | $S_2 = 0$ | FALSE ($S_2$) |
| **3** | $(a \to 0) \equiv (a \to S_1) = S_1'$ | $S_1' = \bar{a}$ |
| **4** | $(b \to S_1') = S_1''$ | $S_1'' = \overline{a.b}$ |
| **5** | $(S_1'' \to 0) \equiv (S_1'' \to S_2) = S_2'$ | $S_2' = a.b = \boldsymbol{AND(a,b)}$ |

Table 7: Serial IMPLY-based implementation of the half adder [28].

| Step | Operation | Equivalent Logic |
|---|---|---|
| **1** | $S_1 = 0$ | FALSE ($S_1$) |
| **2** | $S_2 = 0$ | FALSE ($S_2$) |
| **3** | $(A \to 0) \equiv (A \to S_1) = S_1'$ | $S_1' = \bar{A}$ |
| **4** | $(B \to 0) \equiv (B \to S_2) = S_2'$ | $S_2' = \bar{B}$ |
| **5** | $(S_1' \to S_2') = S_2''$ | $S_2'' = A + \bar{B}$ |
| **6** | $(B \to S_1') = S_1''$ | $S_1'' = \bar{A} + \bar{B}$ |
| **7** | $(A \to B) = B'$ | $B' = \bar{A} + B$ |
| **8** | A = 0 | FALSE (A) |
| **9** | $\left(S_1'' \to 0\right) \equiv \left(S_1'' \to A\right) = A'$ | $A' = A \cdot B = \boldsymbol{C_{out}}$ |
| **10** | $S_1 = 0$ | FALSE ($S_1$) |
| **11** | $\left(S_2'' \to 0\right) \equiv \left(S_2'' \to S_1\right) = S_1'$ | $S_1' = \bar{A} \cdot B$ |
| **12** | $(B' \to S_1') = S_1''$ | $S_1'' = A \cdot \bar{B} + \bar{A} \cdot B = A \oplus B = \boldsymbol{Sum}$ |

First, the implementation algorithms of the AND gate, half adder, and full adder cells in [23, 28] are analyzed separately. The IMPLY-based implementation of the AND gate and the half adder cell [28] are written in (1)–(3), respectively.

$$AND(a,b) = a \cdot b \equiv \left(a \to (b \to 0)\right) \to 0 = \overline{\left(a \to \bar{b}\right)} \tag{1}$$

$$Sum_{HA} = \alpha \oplus \beta \equiv (\alpha \to \beta) \to \overline{\left(\bar{\alpha} \to \bar{\beta}\right)} \tag{2}$$

$$Cout_{HA} = \beta \cdot \alpha \equiv \overline{(\beta \to \bar{\alpha})} \tag{3}$$

The step-by-step IMPLY-based implementation of the AND gate and half adder cell in serial architecture is presented in Tables 6 and 7, respectively. The output of the first four steps, named $\alpha$ here, is inverted in the final step of the AND $(a, b)$ gate's implementation algorithm $(\alpha \to 0)$. By further analysis of the half adder's implementation algorithm introduced in [28], the authors noted that two inputs of the half adder cell (outputs of the two AND gates) are inverted in the third and fourth steps of the algorithm. In short, the $\alpha$ signal of the two AND gates are inverted serially twice: in the last step of the AND gates implementation algorithm and the third and fourth steps of the half adder's algorithm. Hence, it is possible to overlap these steps to reduce the number of computational steps and energy consumption. The proposed $PPU_1$'s output functions are written in (4) and (5).

$$Sum_{PPU1} = ab \oplus cd = \left[\overline{\left(a \to \bar{b}\right)} \to \overline{\left(c \to \bar{d}\right)}\right] \to \overline{\left[\left(a \to \bar{b}\right) \to \left(c \to \bar{d}\right)\right]} \tag{4}$$

$$Cout_{PPU1} = cd \cdot ab \equiv \overline{\left[\overline{(c \rightarrow \bar{d})} \rightarrow (a \rightarrow \bar{b})\right]} \quad (5)$$

Based on these equations, four input and four work memristors are required to implement the proposed IMPLY-based $PPU_1$ serially in eighteen computational steps. The implementation algorithm of the proposed $PPU_1$ is presented in Table 8. The number of computational steps of the proposed $PPU_1$ is reduced by four compared to the classic cell (an improvement of 18%) without extra work memristors.

Table 8: The implementation algorithm of the proposed $PPU_1$.

| Step | Operation | Equivalent Logic |
|---|---|---|
| 1 | $S_1 = 0$ | $FALSE(S_1)$ |
| 2 | $S_2 = 0$ | $FALSE(S_2)$ |
| 3 | $(b \rightarrow 0) \equiv (b \rightarrow S_1) = S_1'$ | $S_1' = \bar{b}$ |
| 4 | $(a \rightarrow S_1') = S_1''$ | $S_1'' = \bar{a} + \bar{b}$ |
| 5 | $(d \rightarrow 0) \equiv (d \rightarrow S_2) = S_2'$ | $S_2' = \bar{d}$ |
| 6 | $(c \rightarrow S_2') = S_2''$ | $S_2'' = \bar{c} + \bar{d}$ |
| 7 | $S_3 = 0$ | $FALSE(S_3)$ |
| 8 | $S_4 = 0$ | $FALSE(S_4)$ |
| 9 | $(S_1'' \rightarrow 0) \equiv (S_1'' \rightarrow S_3) = S_3'$ | $S_3' = ab$ |
| 10 | $(S_2'' \rightarrow 0) \equiv (S_2'' \rightarrow S_4) = S_4'$ | $S_4' = cd$ |
| 11 | $(S_1'' \rightarrow S_2'') = S_2'''$ | $S_2''' = ab + (\bar{c} + \bar{d})$ |
| 12 | $(S_4' \rightarrow S_1'') = S_1'''$ | $S_1''' = (\bar{a} + \bar{b}) + (\bar{c} + \bar{d})$ |
| 13 | $(S_3' \rightarrow S_4') = S_4''$ | $S_4'' = (\bar{a} + \bar{b}) + cd$ |
| 14 | $S_3 = 0$ | $FALSE(S_3)$ |
| 15 | $(S_2''' \rightarrow 0) \equiv (S_2''' \rightarrow S_3) = S_3'$ | $S_3' = (\bar{a} + \bar{b}) \cdot cd$ |
| 16 | $(S_4'' \rightarrow S_3') = S_3''$ | $S_3'' = ab \cdot (\bar{c} + \bar{d}) + (\bar{a} + \bar{b}) \cdot cd = ab \oplus cd$ $= \alpha \oplus \beta = \boldsymbol{Sum}$ |
| 17 | $S_4 = 0$ | $FALSE(S_4)$ |
| 18 | $(S_1''' \rightarrow 0) \equiv (S_1''' \rightarrow S_4) = S_4'$ | $S_4' = ab \cdot cd = \alpha \cdot \beta = \boldsymbol{C_{out}}$ |

The method of overlapping computational steps is applied for other PPUs of the serial unsigned array multiplier, $PPU_2$, and $PPU_3$ as well. As far as the authors know, the serial IMPLY-based full adder proposed in [23] has the minimum hardware complexity and computational steps compared to previous works. The output functions of this full adder cell are written in (6) and (7), and its implementation algorithm is presented in Table 9.

$$Sum_{FA} = x \oplus y \oplus C_{in} \equiv [(\bar{x} \rightarrow y) \rightarrow ((x \rightarrow \bar{y}) \rightarrow C_{in})] \rightarrow \left(\overline{(\overline{x \oplus y}) \rightarrow \overline{C_{in}}}\right) \quad (6)$$

$$Cout_{FA} = x \cdot y + C_{in}(x + y) \equiv \overline{\left[(\bar{x} \rightarrow y) \rightarrow \overline{((x \rightarrow \bar{y}) \rightarrow C_{in})}\right]} \quad (7)$$

The IMPLY operation between the input memristors ($a_{in}$ and $b_{in}$) and two work memristors ($S_1$ and $S_2$) which are both initialized to zero (RESET) is executed in the third and fourth computational steps of the serial full adder's implementation algorithm [23]. In these two computational steps, inputs $a_{in}$ and $b_{in}$ are inverted and temporarily stored in work memristors $S_1$ and $S_2$, as shown in Table 9.

Table 9: Serial IMPLY-based implementation of the full adder [23].

| Step | Operation | Equivalent Logic |
|---|---|---|
| 1 | $S_1 = 0$ | $FALSE(S_1)$ |
| 2 | $S_2 = 0$ | $FALSE(S_2)$ |
| 3 | $(a \rightarrow 0) \equiv (a \rightarrow S_1) = S_1'$ | $S_1' = \bar{a}$ |
| 4 | $(b \rightarrow 0) \equiv (b \rightarrow S_2) = S_2'$ | $S_2' = \bar{b}$ |
| 5 | $(S_1' \rightarrow b) = b'$ | $b' = a + b$ |
| 6 | $(a \rightarrow S_2') = S_2''$ | $S_2'' = \bar{a} + \bar{b}$ |
| 7 | $a = 0$ | $FALSE(a)$ |
| 8 | $(b' \rightarrow 0) \equiv (b' \rightarrow a) = a'$ | $a' = \bar{a} \cdot \bar{b}$ |
| 9 | $(S_2'' \rightarrow a') = a''$ | $a'' = ab + \bar{a}\bar{b} = \overline{a \oplus b}$ |
| 10 | $S_1 = 0$ | $FALSE(S_1)$ |
| 11 | $(C_{in} \rightarrow 0) \equiv (C_{in} \rightarrow S_1) = S_1'$ | $S_1' = \overline{C_{in}}$ |
| 12 | $(S_2'' \rightarrow C_{in}) = C_{in}'$ | $C_{in}' = ab + C_{in}$ |
| 13 | $(a'' \rightarrow S_1') = S_1''$ | $S_1'' = (a \oplus b) + \overline{C_{in}}$ |
| 14 | $a = 0$ | $FALSE(a)$ |
| 15 | $(S_1'' \rightarrow 0) \equiv (S_1'' \rightarrow a) = a'$ | $a' = \overline{a \oplus b} \cdot C_{in}$ |
| 16 | $S_2 = 0$ | $FALSE(S_2)$ |
| 17 | $(C_{in}' \rightarrow 0) \equiv (C_{in}' \rightarrow S_2) = S_2'$ | $S_2' = \left(\bar{a} + \bar{b}\right) \cdot \overline{C_{in}}$ |
| 18 | $(b' \rightarrow S_2') = S_2''$ | $S_2'' = \bar{a}\bar{b} + \left(\bar{a} + \bar{b}\right) \cdot \overline{C_{in}}$ |
| 19 | $(b' \rightarrow C_{in}') = C_{in}''$ | $C_{in}'' = \bar{a}\bar{b} + (ab + C_{in})$ |
| 20 | $(C_{in}'' \rightarrow a') = a''$ | $a'' = \left[(a+b) \cdot \left(\left(\bar{a} + \bar{b}\right) \cdot \overline{C_{in}}\right)\right] + \left(\overline{a \oplus b} \cdot C_{in}\right)$ $= (a \oplus b \cdot \overline{C_{in}}) + \left(\overline{a \oplus b} \cdot C_{in}\right) = a \oplus b \oplus C_{in} = \boldsymbol{Sum}$ |
| 21 | $C_{in} = 0$ | $FALSE(C_{in})$ |
| 22 | $(S_2'' \rightarrow C_{in}) = C_{in}'$ | $C_{in}' = (a+b) \cdot (ab + C_{in}) = ab + C_{in}(a+b) = \boldsymbol{C_{out}}$ |

In $PPU_2$, one of the two inputs of the full adder cell is fed by the AND gate, and in $PPU_3$, both of these inputs are fed by the output of the AND gates. As mentioned earlier, α is inverted in the last computational step of the AND gate's implementation algorithm $(\alpha \rightarrow 0)$. Therefore, it is possible to overlap the last computational step of the AND gate (two AND gates) in $PPU_2$ ($PPU_3$) with the third (third and fourth) computational step of the full adder cell implementation algorithm proposed in [23]. As a result, the computational steps of the proposed $PPU_2$ cell (25 steps) are improved by 7% compared to the classic design (27 steps). Moreover, the number of computational steps of the proposed $PPU_3$ (28 steps) is reduced by 12.5% compared to the classic $PPU_3$ (32 steps). This reduction of computational steps is made possible by overlapping repeated serial steps of the subcomponents of these PPUs without increasing the number of work memristors. The output equations of $PPU_2$ and $PPU_3$ are written in (8)–(11), respectively.

$$\mathbf{PPU_2}: Sum_{FA} = ab \oplus \beta \oplus C_{in} \equiv \left[\left(\left(\overline{(a \rightarrow \bar{b})} \rightarrow \beta\right) \rightarrow \overline{\left((a \rightarrow \bar{b}) \rightarrow \bar{\beta}\right)}\right) \rightarrow C_{in}\right] \rightarrow \overline{\left[\left(\overline{\left(\left(\overline{(a \rightarrow \bar{b})} \rightarrow \beta\right) \rightarrow \overline{\left((a \rightarrow \bar{b}) \rightarrow \bar{\beta}\right)}\right)} \rightarrow \overline{C_{in}}\right)\right]} \quad (8)$$

$$Cout_{FA} = \beta \cdot ab + C_{in}(ab + \beta) \equiv \left(\beta \to \left(a \to \bar{b}\right)\right) \to \overline{\left(C_{in} \to \overline{\left(\left(a \to \bar{b}\right) \to \beta\right)}\right)} \tag{9}$$

$$\textbf{PPU}_\mathbf{3}: Sum_{FA} = ab \oplus cd \oplus C_{in} \equiv \left[\left(\left(\overline{\left(a \to \bar{b}\right)} \to \overline{\left(c \to \bar{d}\right)}\right) \to \overline{\left(\left(a \to \bar{b}\right) \to \left(c \to \bar{d}\right)\right)}\right) \to C_{in}\right]$$
$$\to \overline{\left[\left(\left(\overline{\left(a \to \bar{b}\right)} \to \overline{\left(c \to \bar{d}\right)}\right) \to \overline{\left(\left(a \to \bar{b}\right) \to \left(c \to \bar{d}\right)\right)}\right) \to \overline{C_{in}}\right]} \tag{10}$$

$$Cout_{FA} = cd \cdot ab + C_{in}(ab + cd) \equiv \left(\overline{\left(c \to \bar{d}\right)} \to \left(a \to \bar{b}\right)\right) \to \overline{\left(C_{in} \to \overline{\left(\left(a \to \bar{b}\right) \to \overline{\left(c \to \bar{d}\right)}\right)}\right)} \tag{11}$$

The implementation algorithm of *$PPU_2$* and *$PPU_3$* are tabulated in Tables 10 and 11, respectively.

Table 10: The implementation algorithm of the proposed *$PPU_2$*.

| Step | Operation | Equivalent Logic |
|---|---|---|
| **1** | $S_1 = 0$ | $FALSE(S_1)$ |
| **2** | $S_2 = 0$ | $FALSE(S_2)$ |
| **3** | $S_3 = 0$ | $FALSE(S_3)$ |
| **4** | $(b \to 0) \equiv (b \to S_1) = S_1{'}$ | $S_1{'} = \bar{b}$ |
| **5** | $(a \to S_1{'}) = S_1{''}$ | $S_1{''} = \bar{a} + \bar{b}$ |
| **6** | $(S_1{''} \to 0) \equiv (S_1{''} \to S_3) = S_3{'}$ | $S_3{'} = ab = \alpha$ |
| **7** | $(\beta \to 0) \equiv (\beta \to S_2) = S_2{'}$ | $S_2{'} = \bar{\beta}$ |
| **8** | $(S_1{''} \to \beta) = \beta'$ | $\beta' = ab + \beta$ |
| **9** | $(S_3{'} \to S_2{'}) = S_2{''}$ | $S_2{''} = \left(\bar{a} + \bar{b}\right) + \bar{\beta}$ |
| **10** | $S_3 = 0$ | $FALSE(S_3)$ |
| **11** | $(\beta' \to 0) \equiv (\beta' \to S_3) = S_3{'}$ | $S_3{'} = \left(\bar{a} + \bar{b}\right) \cdot \bar{\beta}$ |
| **12** | $(S_2{''} \to S_3{'}) = S_3{''}$ | $S_3{''} = ab \cdot \beta + \left(\bar{a} + \bar{b}\right) \cdot \bar{\beta} = \overline{\alpha \oplus \beta}$ |
| **13** | $S_1 = 0$ | $FALSE(S_1)$ |
| **14** | $(C_{in} \to 0) \equiv (C_{in} \to S_1) = S_1{'}$ | $S_1{'} = \overline{C_{in}}$ |
| **15** | $(S_2{''} \to C_{in}) = C_{in}{'}$ | $C_{in}{'} = ab \cdot \beta + C_{in}$ |
| **16** | $(S_3{''} \to S_1{'}) = S_1{''}$ | $S_1{''} = (\alpha \oplus \beta) + \overline{C_{in}}$ |
| **17** | $S_3 = 0$ | $FALSE(S_3)$ |
| **18** | $(S_1{''} \to 0) \equiv (S_1{''} \to S_3) = S_3{'}$ | $S_3{'} = \overline{\alpha \oplus \beta} \cdot C_{in}$ |
| **19** | $S_1 = 0$ | $FALSE(S_1)$ |
| **20** | $(C_{in}{'} \to 0) \equiv (C_{in}{'} \to S_1) = S_1{'}$ | $S_1{'} = \left(\left(\bar{a} + \bar{b}\right) + \bar{\beta}\right) \cdot \overline{C_{in}}$ |
| **21** | $(\beta' \to S_1{'}) = S_1{''}$ | $S_1{''} = \left(\bar{a} + \bar{b}\right) \cdot \bar{\beta} + \left(\left(\bar{a} + \bar{b}\right) + \bar{\beta}\right) \cdot \overline{C_{in}}$ |
| **22** | $(\beta' \to C_{in}{'}) = C_{in}{''}$ | $C_{in}{''} = \left(\bar{a} + \bar{b}\right) \cdot \bar{\beta} + \ (ab \cdot \beta + C_{in})$ |
| **23** | $(C_{in}{''} \to S_3{'}) = S_3{''}$ | $S_3{''} = \left[(\alpha + \beta) \cdot \left(\left(\bar{\alpha} + \bar{\beta}\right) \cdot \overline{C_{in}}\right)\right] + \left(\overline{\alpha \oplus \beta} \cdot C_{in}\right)$ <br> $= (\alpha \oplus \beta \cdot \overline{C_{in}}) + \left(\overline{\alpha \oplus \beta} \cdot C_{in}\right) = \alpha \oplus \beta \oplus C_{in} = \boldsymbol{Sum}$ |
| **24** | $S_2 = 0$ | $FALSE(S_2)$ |
| **25** | $(S_1{''} \to S_2) = S_2{'}$ | $S_2{'} = (\alpha + \beta) \cdot (\alpha \cdot \beta + C_{in}) = \alpha \cdot \beta + C_{in}(\alpha + \beta) = \boldsymbol{C_{out}}$ |

Table 11: The implementation algorithm of the proposed $PPU_3$.

| Step | Operation | Equivalent Logic |
|---|---|---|
| 1 | $S_1 = 0$ | $FALSE(S_1)$ |
| 2 | $S_2 = 0$ | $FALSE(S_2)$ |
| 3 | $(b \rightarrow 0) \equiv (b \rightarrow S_1) = S_1'$ | $S_1' = \bar{b}$ |
| 4 | $(a \rightarrow S_1') = S_1''$ | $S_1'' = \bar{a} + \bar{b}$ |
| 5 | $(d \rightarrow 0) \equiv (d \rightarrow S_2) = S_2'$ | $S_2' = \bar{d}$ |
| 6 | $(c \rightarrow S_2') = S_2''$ | $S_2'' = \bar{c} + \bar{d}$ |
| 7 | $S_3 = 0$ | $FALSE(S_3)$ |
| 8 | $S_4 = 0$ | $FALSE(S_4)$ |
| 9 | $(S_1'' \rightarrow 0) \equiv (S_1'' \rightarrow S_3) = S_3'$ | $S_3' = ab = \alpha$ |
| 10 | $(S_2'' \rightarrow 0) \equiv (S_2'' \rightarrow S_4) = S_4'$ | $S_4' = cd = \beta$ |
| 11 | $(S_1'' \rightarrow S_4') = S_4''$ | $S_4'' = ab + cd$ |
| 12 | $(S_3' \rightarrow S_2'') = S_2'''$ | $S_2''' = (\bar{a} + \bar{b}) + (\bar{c} + \bar{d})$ |
| 13 | $S_3 = 0$ | $FALSE(S_3)$ |
| 14 | $(S_4'' \rightarrow 0) \equiv (S_4'' \rightarrow S_3) = S_3'$ | $S_3' = (\bar{a} + \bar{b}) \cdot (\bar{c} + \bar{d})$ |
| 15 | $(S_2''' \rightarrow S_3') = S_3''$ | $S_3'' = ab \cdot cd + (\bar{a} + \bar{b}) \cdot (\bar{c} + \bar{d}) = \overline{\alpha \oplus \beta}$ |
| 16 | $S_1 = 0$ | $FALSE(S_1)$ |
| 17 | $(C_{in} \rightarrow 0) \equiv (C_{in} \rightarrow S_1) = S_1'$ | $S_1' = \overline{C_{in}}$ |
| 18 | $(S_2''' \rightarrow C_{in}) = C_{in}'$ | $C_{in}' = ab \cdot cd + C_{in}$ |
| 19 | $(S_3'' \rightarrow S_1') = S_1''$ | $S_1'' = (\alpha \oplus \beta) + \overline{C_{in}}$ |
| 20 | $S_3 = 0$ | $FALSE(S_3)$ |
| 21 | $(S_1'' \rightarrow 0) \equiv (S_1'' \rightarrow S_3) = S_3'$ | $S_3' = \overline{\alpha \oplus \beta} \cdot C_{in}$ |
| 22 | $S_2 = 0$ | $FALSE(S_2)$ |
| 23 | $(C_{in}' \rightarrow 0) \equiv (C_{in}' \rightarrow S_2) = S_2'$ | $S_2' = \left((\bar{a} + \bar{b}) + (\bar{c} + \bar{d})\right) \cdot \overline{C_{in}}$ |
| 24 | $(S_4'' \rightarrow S_2') = S_2''$ | $S_2'' = (\bar{a} + \bar{b}) \cdot (\bar{c} + \bar{d}) + \left((\bar{a} + \bar{b}) + (\bar{c} + \bar{d})\right) \cdot \overline{C_{in}}$ |
| 25 | $(S_4'' \rightarrow C_{in}') = C_{in}''$ | $C_{in}'' = (\bar{a} + \bar{b}) \cdot (\bar{c} + \bar{d}) + (ab \cdot cd + C_{in})$ |
| 26 | $(C_{in}'' \rightarrow S_3') = S_3''$ | $S_3'' = \left[(\alpha + \beta) \cdot \left((\bar{\alpha} + \bar{\beta}) \cdot \overline{C_{in}}\right)\right] + \left(\overline{\alpha \oplus \beta} \cdot C_{in}\right)$ $= (\alpha \oplus \beta \cdot \overline{C_{in}}) + \left(\overline{\alpha \oplus \beta} \cdot C_{in}\right) = \alpha \oplus \beta \oplus C_{in} = \boldsymbol{Sum}$ |
| 27 | $S_4 = 0$ | $FALSE(S_4)$ |
| 28 | $(S_2'' \rightarrow S_4) = S_4'$ | $S_4' = (\alpha + \beta) \cdot (\alpha \cdot \beta + C_{in}) = \alpha \cdot \beta + C_{in}(\alpha + \beta) = \boldsymbol{C_{out}}$ |

The computational steps of the unsigned array multiplier based on the proposed PPUs' implementation algorithms are reduced compared to the classic structure. The total number of computational steps of the proposed n-bit unsigned CSA-based array multiplier is $25n^2 - 32n + 2$ which is obtained from (12). $CS_{UAmultiplier}$ refers to the number of computational steps of the unsigned array multiplier in (12). Moreover, *5n – 4* memristors (*2n* input and *3n – 4* work memristors) are needed to implement the proposed unsigned array multiplier.

$$CS_{UAmultiplier} = \#PPU_1 \times CS_{PPU1} + \#PPU_2 \times CS_{PPU2} + \#PPU_3 \times CS_{PPU3} + \#HA \times CS_{HA} + \#FA \times CS_{FA} + \#AND \times CS_{AND} \quad (12)$$

### 3.2 Proposed signed array multiplier

CSA-based array multiplier can also be applied to calculate the product of signed numbers. In an n-bit signed CSA-based array multiplier, PPUs are placed in columns *1* to $2n + 1$ (in $n - 1$ diagonal lines) and $n$ rows. Figure 3 depicts the structure of a 4-bit signed array multiplier based on CSA. The main difference between signed and unsigned CSA-based array multipliers lies in their types of PPUs. As explained earlier, the unsigned CSA-based array multiplier consists of three types of PPUs, but the signed array multiplier comprises eight different PPU structures which are indicated with different colors in Figure 3.

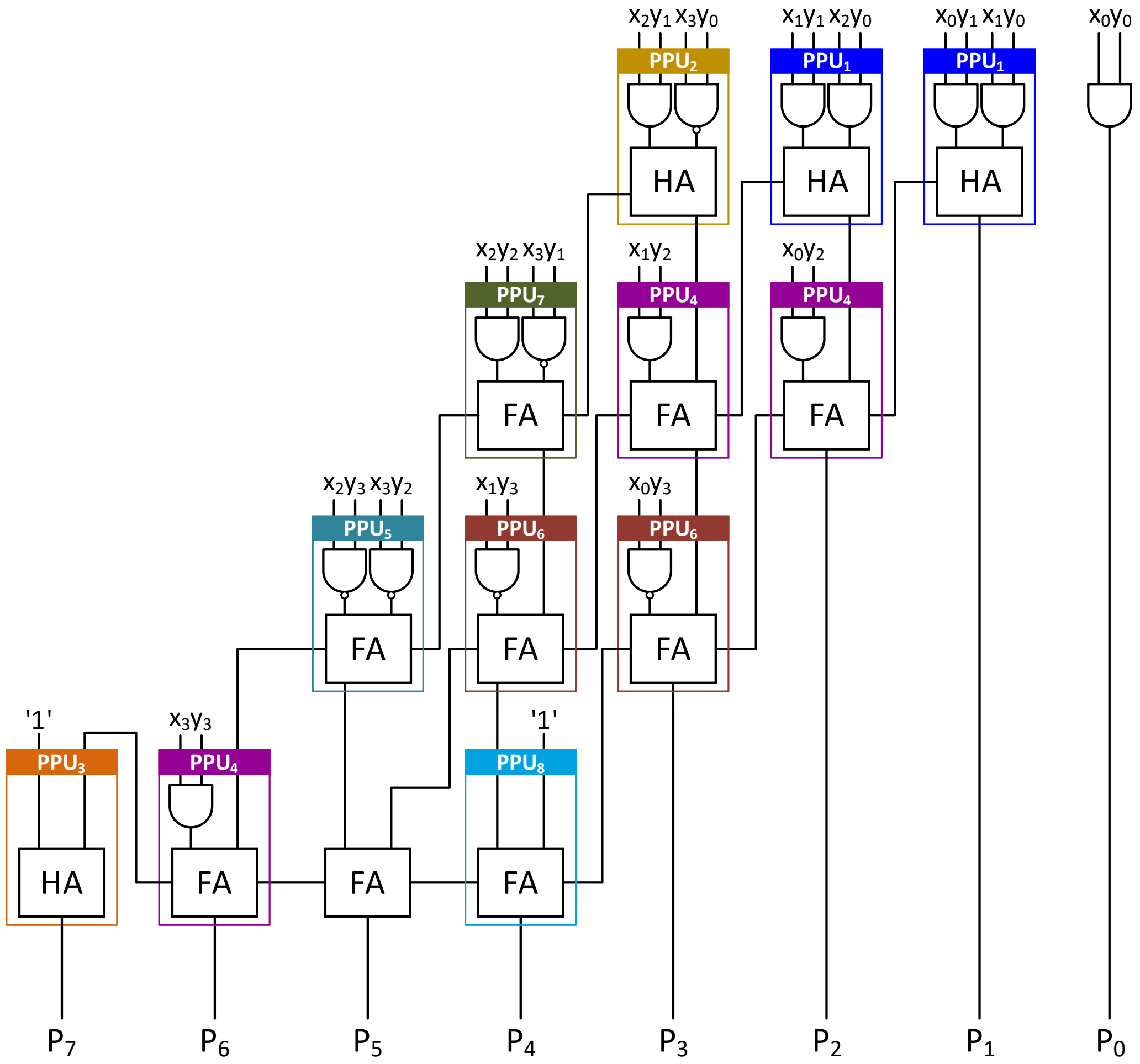

Figure 3: 4-bit signed CSA-based array multiplier.

In the signed CSA-based array multiplier, AND and NAND gates are applied to generate partial products (PPG stage). Half adder and full adder cells are also used to reduce the multiplication tree (PPR stage) and final addition (RCA stage). Eight types of PPUs form the signed CSA-based array multiplier containing different combinations of these logic gates and arithmetic cells. $PPU_1$ of the signed CSA-based array multiplier has the same structure as $PPU_1$ of the unsigned multiplier in which two inputs of the half adder cell are the outputs of two AND gates. In the structure of $PPU_2$, one of the inputs of the half adder cell is connected to the output of a NAND gate and the other to the output of an AND gate. There is only one $PPU_3$ cell in the structure of the signed array multiplier, which is a half adder located in the $2n^{th}$ column. One of the inputs of this PPU is a constant bit '1', and its other input is the $C_{out}$ of the full adder cell located in the $n^{th}$ row and $(2n - 1)^{th}$ column. In the $PPU_4$–$PPU_8$ structure, the full adder cell is applied to reduce the partial products and compute the final results in the RCA stage. In these five PPUs, one of the inputs of the full adder cell, located in the $i^{th}$ row and $j^{th}$ column, is the $C_{out}$ of the PPU located in the $(i - 1)^{th}$ row and $(j - 1)^{th}$ column. In $PPU_4$ and $PPU_6$, one of the inputs of the full adder cell (located in the $i^{th}$ row) is the *Sum* output of the half adder or full adder cell of the PPU placed in the $(i-1)^{th}$ row. The other input of $PPU_4$/$PPU_6$ is the AND/NAND gate output. In the structure of $PPU_5$, the two inputs of the full adder are fed from the partial products generated by two NAND gates. This PPU is applied in the $(n-1)^{th}$ row and $(2n-2)^{th}$ column. $PPU_7$ cells are placed in the last column of $2^{nd}$ to $(n-2)^{th}$ rows, in which two inputs result from partial products generated by an AND gate and a NAND gate. The last PPU, $PPU_8$, is located in $\left(\frac{n}{2}+1\right)^{th}$ column and $n^{th}$ row of an n-bit signed array multiplier. One of this PPU's inputs is the constant bit '1' [27], and the other input is the *Sum* output of the full adder ($PPU_4$), which is located in $(n-1)^{th}$ row and $\left(\frac{n}{2}+1\right)^{th}$ column. Table 12 lists the number of applied $PPU_1$–$PPU_8$ cells and their computational steps in the proposed n-bit signed CSA-based array multiplier.

Table 12: Characteristics of the proposed CSA-based signed array multiplier.

| Combinational Block | No. of Applied Blocks | No. of Computational Steps |
|---|---|---|
| $PPU_1$ | $n-2$ | 18 |
| $PPU_2$ | 1 | 18 |
| $PPU_3$ | 1 | 2 |
| $PPU_4$ | $n^2-5n+7$ | 25 |
| $PPU_5$ | 1 | 28 |
| $PPU_6$ | $n-2$ | 25 |
| $PPU_7$ | $n-3$ | 28 |
| $PPU_8$ | 1 | 9 |
| Full Adder [23] | $n-3$ | 22 |
| AND gate | 1 | 5 |

Similar to the serial unsigned array multiplier, the method of overlapping repetitive serial steps has been applied to reduce the computational steps and energy consumption of the n-bit signed CSA-based array multiplier in serial architecture. The structures of $PPU_3$ and $PPU_8$, containing one constant input with a value of '1', have also been redefined to reduce these criteria further. $PPU_1$ and $PPU_4$ of the n-bit signed array multiplier are the same as the unsigned array multiplier's $PPU_1$ and $PPU_2$, respectively. Hence, repeated serial steps can be overlapped in the signed array multiplier's $PPU_1$ and $PPU_4$, the same as the ones in the unsigned array multiplier. The number of computational steps required to implement $PPU_1$ and $PPU_4$ in the serial signed CSA-based array multiplier is reduced to 18 (18% improvement) and 25 (7% improvement), respectively. The implementation algorithm of PPUs' constituent building blocks, including NAND gate (see Table 13), AND gate (see (1) and Table 6), half adder (see (2), (3), and Table 7) and full adder (see (6) and (7) and Table 9) has been investigated in detail to apply the overlapping repetitive serial steps method to reduce the computational steps of each PPU, resulting in a reduction of computational steps and energy consumption of the proposed signed array multiplier.

Table 13: Serial IMPLY-based implementation of the NAND gate.

| Step | Operation | Equivalent Logic |
|---|---|---|
| **1** | $S_1 = 0$ | FALSE ($S_1$) |
| **2** | $(b \rightarrow 0) \equiv (b \rightarrow S_1) = S_1'$ | $S_1' = \bar{b}$ |
| **3** | $(a \rightarrow S_1') = S_1''$ | $S_1'' = \overline{a.b} = \boldsymbol{NAND(a,b)}$ |

The NAND gate's output is computed in three computational steps and stored in a work memristor based on its serial implementation algorithm. The NAND gate's output (the half adder and full adder cells' input in $PPU_{2,5,6,7}$) is inverted once, according to (2), (3), (6), and (7) and Tables 7 and 9. Hence, there are no repetitive serial steps to be overlapped between the NAND gate and the half adder and full adder cells. Conversely, the last two computational steps of the AND gate are repeated as the first two steps of the half adder and full adder's algorithm and are overlapped in $PPU_2$ and $PPU_7$. The proposed $PPU_2$ and $PPU_7$ implementation algorithms are shown in Tables 14 and 15, respectively, based on (13)–(16).

$$\mathbf{PPU_2}:\ Sum_{HA} = ab\oplus\overline{cd} \equiv \left[\overline{(a\to\bar{b})}\to(c\to\bar{d})\right]\to\overline{\left[(a\to\bar{b})\to\overline{(c\to\bar{d})}\right]} \quad (13)$$

$$Cout_{HA} = ab\cdot\overline{cd} \equiv \overline{\left[(c\to\bar{d})\to(a\to\bar{b})\right]} \quad (14)$$

$$\mathbf{PPU_7}:\ Sum_{FA} = \overline{ab}\oplus cd\oplus C_{in} \equiv \left[\left(\overline{(a\to\bar{b})}\to\overline{(c\to\bar{d})}\right)\to\left(\left((a\to\bar{b})\to(c\to\bar{d})\right)\to C_{in}\right)\right]$$
$$\to\overline{\left[\left(\left(\overline{(a\to\bar{b})}\to\overline{(c\to\bar{d})}\right)\to\overline{\left((a\to\bar{b})\to(c\to\bar{d})\right)}\right)\to\overline{C_{in}}\right]} \quad (15)$$

$$Cout_{FA} = \overline{ab}\cdot cd + C_{in}\left(\overline{ab}+cd\right)$$
$$\equiv\overline{\left[\left(\overline{(a\to\bar{b})}\to\overline{(c\to\bar{d})}\right)\to\left(\overline{\left(\left((a\to\bar{b})\to(c\to\bar{d})\right)\to C_{in}\right)}\right)\right]} \quad (16)$$

The computational steps of the proposed $PPU_2$ and $PPU_7$ equal 18 steps (10% improvement) and 28 steps (7% improvement), respectively, reduced by two steps compared to their classic circuits (20 and 30 steps). The number of memristors required for these two proposed PPUs has not changed compared to the classic structures.

Table 14: The implementation algorithm of the proposed $PPU_2$ in the signed array multiplier.

| Step | Operation | Equivalent Logic |
|---|---|---|
| 1 | $S_1 = 0$ | $FALSE(S_1)$ |
| 2 | $S_2 = 0$ | $FALSE(S_2)$ |
| 3 | $(b\to 0)\equiv(b\to S_1)=S_1'$ | $S_1' = \bar{b}$ |
| 4 | $(a\to S_1')=S_1''$ | $S_1'' = \bar{a}+\bar{b}=\bar{\alpha}$ |
| 5 | $(d\to 0)\equiv(d\to S_2)=S_2'$ | $S_2' = \bar{d}$ |
| 6 | $(c\to S_2')=S_2''$ | $S_2'' = \bar{c}+\bar{d}=\beta$ |
| 7 | $S_3 = 0$ | $FALSE(S_3)$ |
| 8 | $S_4 = 0$ | $FALSE(S_4)$ |
| 9 | $(S_1''\to 0)\equiv(S_1''\to S_3)=S_3'$ | $S_3' = ab = \alpha$ |
| 10 | $(S_2''\to 0)\equiv(S_2''\to S_4)=S_4'$ | $S_4' = cd = \bar{\beta}$ |
| 11 | $(S_1''\to S_4')=S_4''$ | $S_4'' = ab+cd=\alpha+\bar{\beta}$ |
| 12 | $(S_2''\to S_1'')=S_1'''$ | $S_1''' = cd+\left(\bar{a}+\bar{b}\right)=\bar{\beta}+\bar{\alpha}$ |
| 13 | $(S_3'\to S_2'')=S_2'''$ | $S_2''' = \left(\bar{a}+\bar{b}\right)+\left(\bar{c}+\bar{d}\right)=\bar{\alpha}+\beta$ |
| 14 | $S_3 = 0$ | $FALSE(S_3)$ |
| 15 | $(S_4''\to 0)\equiv(S_4''\to S_3)=S_3'$ | $S_3' = \left(\bar{a}+\bar{b}\right)\cdot\left(\bar{c}+\bar{d}\right)=\bar{\alpha}\cdot\beta$ |
| 16 | $(S_2'''\to S_3')=S_3''$ | $S_3'' = (ab\cdot cd)+\left(\left(\bar{a}+\bar{b}\right)\cdot\left(\bar{c}+\bar{d}\right)\right) = \alpha\oplus\beta = \boldsymbol{Sum}$ |
| 17 | $S_2 = 0$ | $FALSE(S_2)$ |
| 18 | $(S_1'''\to 0)\equiv(S_1'''\to S_2)=S_2'$ | $S_2' = \left(\bar{c}+\bar{d}\right)\cdot ab=\beta\cdot\alpha=\boldsymbol{C_{out}}$ |

Table 15: The implementation algorithm of the proposed *$PPU_7$* in the signed array multiplier.

| Step | Operation | Equivalent Logic |
|---|---|---|
| 1 | $S_1 = 0$ | $FALSE(S_1)$ |
| 2 | $S_2 = 0$ | $FALSE(S_2)$ |
| 3 | $S_3 = 0$ | $FALSE(S_3)$ |
| 4 | $S_4 = 0$ | $FALSE(S_4)$ |
| 5 | $(b \to 0) \equiv (b \to S_1) = S_1'$ | $S_1' = \bar{b}$ |
| 6 | $(a \to S_1') = S_1''$ | $S_1'' = \bar{a} + \bar{b} = \alpha$ |
| 7 | $(d \to 0) \equiv (d \to S_2) = S_2'$ | $S_2' = \bar{d}$ |
| 8 | $(c \to S_2') = S_2''$ | $S_2'' = \bar{c} + \bar{d} = \bar{\beta}$ |
| 9 | $(S_1'' \to 0) \equiv (S_1'' \to S_3) = S_3'$ | $S_3' = ab = \bar{\alpha}$ |
| 10 | $(S_2'' \to 0) \equiv (S_2'' \to S_4) = S_4'$ | $S_4' = cd = \beta$ |
| 11 | $(S_3' \to S_4') = S_4''$ | $S_4'' = \left(\bar{a} + \bar{b}\right) + cd = \alpha + \beta$ |
| 12 | $(S_1'' \to S_2'') = S_2'''$ | $S_2''' = ab + \left(\bar{c} + \bar{d}\right) = \bar{\alpha} + \bar{\beta}$ |
| 13 | $S_1 = 0$ | $FALSE(S_1)$ |
| 14 | $(S_2''' \to 0) \equiv (S_2''' \to S_1) = S_1'$ | $S_1' = \left(\bar{a} + \bar{b}\right) \cdot cd = \alpha \cdot \beta$ |
| 15 | $(S_4'' \to S_1') = S_1''$ | $S_1'' = ab \cdot \left(\bar{c} + \bar{d}\right) + \left(\bar{a} + \bar{b}\right) \cdot cd = \bar{\alpha} \cdot \bar{\beta} + \alpha \cdot \beta$ |
| 16 | $S_3 = 0$ | $FALSE(S_3)$ |
| 17 | $(C_{in} \to 0) \equiv (C_{in} \to S_3) = S_3'$ | $S_3' = \overline{C_{in}}$ |
| 18 | $(S_1'' \to S_3') = S_3''$ | $S_3'' = \left(\left(\bar{a} + \bar{b}\right) + cd\right) \cdot \left(ab + \left(\bar{c} + \bar{d}\right)\right) + \overline{C_{in}}$<br>$= (\alpha + \beta) \cdot \left(\bar{\alpha} + \bar{\beta}\right) + \overline{C_{in}}$ |
| 19 | $S_1 = 0$ | $FALSE(S_1)$ |
| 20 | $(S_3'' \to 0) \equiv (S_3'' \to S_1) = S_1'$ | $S_1' = \left(ab \cdot \left(\bar{c} + \bar{d}\right) + \left(\bar{a} + \bar{b}\right) \cdot cd\right) \cdot C_{in}$<br>$= \left(\bar{\alpha} \cdot \bar{\beta} + \alpha \cdot \beta\right) \cdot C_{in}$ |
| 21 | $(S_2''' \to C_{in}) = C_{in}'$ | $C_{in}' = \left(\bar{a} + \bar{b}\right) \cdot cd + C_{in} = \alpha \cdot \beta + C_{in}$ |
| 22 | $S_2 = 0$ | $FALSE(S_2)$ |
| 23 | $(C_{in}' \to 0) \equiv (C_{in}' \to S_2) = S_2'$ | $S_2' = \left(ab + \left(\bar{c} + \bar{d}\right)\right) \cdot \overline{C_{in}} = \left(\bar{\alpha} + \bar{\beta}\right) \cdot \overline{C_{in}}$ |
| 24 | $(S_4'' \to S_2') = S_2''$ | $S_2'' = ab \cdot \left(\bar{c} + \bar{d}\right) + \left(\left(ab + \left(\bar{c} + \bar{d}\right)\right) \cdot \overline{C_{in}}\right)$<br>$= \bar{\alpha} \cdot \bar{\beta} + \left(\left(\bar{\alpha} + \bar{\beta}\right) \cdot \overline{C_{in}}\right)$ |
| 25 | $(S_4'' \to C_{in}') = C_{in}''$ | $C_{in}'' = ab \cdot \left(\bar{c} + \bar{d}\right) + \left(\left(\bar{a} + \bar{b}\right) \cdot cd + C_{in}\right)$<br>$= \bar{\alpha} \cdot \bar{\beta} + (\alpha \cdot \beta + C_{in})$ |
| 26 | $(C_{in}'' \to S_1') = S_1''$ | $S_1'' = \left[(\alpha + \beta) \cdot \left(\left(\bar{\alpha} + \bar{\beta}\right) \cdot \overline{C_{in}}\right)\right] + \left(\overline{\alpha \oplus \beta} \cdot C_{in}\right)$<br>$= \alpha \oplus \beta \oplus C_{in} = \boldsymbol{Sum}$ |
| 27 | $S_3 = 0$ | $FALSE(S_3)$ |
| 28 | $(S_2'' \to S_3) = S_3'$ | $S_3' = (\alpha + \beta) \cdot (\alpha \cdot \beta + C_{in})$<br>$= \alpha \cdot \beta + C_{in}(\alpha + \beta) = \boldsymbol{C_{out}}$ |

Table 16: The implementation algorithm of the proposed *PPU*$_5$ in the signed array multiplier.

| Step | Operation | Equivalent Logic |
|---|---|---|
| 1 | $S_1 = 0$ | $FALSE(S_1)$ |
| 2 | $S_2 = 0$ | $FALSE(S_2)$ |
| 3 | $S_3 = 0$ | $FALSE(S_3)$ |
| 4 | $S_4 = 0$ | $FALSE(S_4)$ |
| 5 | $(b \rightarrow 0) \equiv (b \rightarrow S_1) = S_1'$ | $S_1' = \bar{b}$ |
| 6 | $(a \rightarrow S_1') = S_1''$ | $S_1'' = \bar{a} + \bar{b} = \alpha$ |
| 7 | $(d \rightarrow 0) \equiv (d \rightarrow S_2) = S_2'$ | $S_2' = \bar{d}$ |
| 8 | $(c \rightarrow S_2') = S_2''$ | $S_2'' = \bar{c} + \bar{d} = \beta$ |
| 9 | $(S_1'' \rightarrow 0) \equiv (S_1'' \rightarrow S_3) = S_3'$ | $S_3' = ab = \bar{\alpha}$ |
| 10 | $(S_2'' \rightarrow 0) \equiv (S_2'' \rightarrow S_4) = S_4'$ | $S_4' = cd = \bar{\beta}$ |
| 11 | $(S_3' \rightarrow S_2'') = S_2'''$ | $S_2''' = (\bar{a} + \bar{b}) + (\bar{c} + \bar{d})$ |
| 12 | $(S_1'' \rightarrow S_4') = S_4''$ | $S_4'' = ab + cd$ |
| 13 | $S_1 = 0$ | $FALSE(S_1)$ |
| 14 | $(S_2''' \rightarrow 0) \equiv (S_2''' \rightarrow S_1) = S_1'$ | $S_1' = ab \cdot cd$ |
| 15 | $(S_4'' \rightarrow S_1') = S_1''$ | $S_1'' = (\bar{a} + \bar{b}) \cdot (\bar{c} + \bar{d}) + ab \cdot cd = \overline{\alpha \oplus \beta}$ |
| 16 | $S_3 = 0$ | $FALSE(S_3)$ |
| 17 | $(C_{in} \rightarrow 0) \equiv (C_{in} \rightarrow S_3) = S_3'$ | $S_3' = \overline{C_{in}}$ |
| 18 | $(S_4'' \rightarrow C_{in}) = C_{in}'$ | $C_{in}' = (\bar{a} + \bar{b}) \cdot (\bar{c} + \bar{d}) + C_{in}$ |
| 19 | $(S_1'' \rightarrow S_3') = S_3''$ | $S_3'' = (\alpha \oplus \beta) + \overline{C_{in}}$ |
| 20 | $S_4 = 0$ | $FALSE(S_4)$ |
| 21 | $(S_3'' \rightarrow 0) \equiv (S_3'' \rightarrow S_4) = S_4'$ | $S_4' = \overline{\alpha \oplus \beta} \cdot C_{in}$ |
| 22 | $S_1 = 0$ | $FALSE(S_1)$ |
| 23 | $(C_{in}' \rightarrow 0) \equiv (C_{in}' \rightarrow S_1) = S_1'$ | $S_1' = (ab + cd) \cdot \overline{C_{in}}$ |
| 24 | $(S_2''' \rightarrow S_1') = S_1''$ | $S_1'' = ab \cdot cd + (ab + cd) \cdot \overline{C_{in}}$<br>$= \bar{\alpha} \cdot \bar{\beta} + (\bar{\alpha} + \bar{\beta}) \cdot \overline{C_{in}}$ |
| 25 | $(S_2''' \rightarrow C_{in}') = C_{in}''$ | $C_{in}'' = ab \cdot cd + \left((\bar{a} + \bar{b}) \cdot (\bar{c} + \bar{d}) + C_{in}\right)$<br>$= \bar{\alpha} \cdot \bar{\beta} + (\alpha \cdot \beta + C_{in})$ |
| 26 | $(C_{in}'' \rightarrow S_4') = S_4''$ | $S_4'' = \left[(\alpha + \beta) \cdot \left((\bar{\alpha} + \bar{\beta}) \cdot \overline{C_{in}}\right)\right] + \left(\overline{\alpha \oplus \beta} \cdot C_{in}\right)$<br>$= (\alpha \oplus \beta \cdot \overline{C_{in}}) + \left(\overline{\alpha \oplus \beta} \cdot C_{in}\right)$<br>$= \alpha \oplus \beta \oplus C_{in} = \boldsymbol{Sum}$ |
| 27 | $S_3 = 0$ | $FALSE(S_3)$ |
| 28 | $(S_1'' \rightarrow 0) \equiv (S_1'' \rightarrow S_3) = S_3'$ | $S_3' = (\alpha + \beta) \cdot (\alpha \cdot \beta + C_{in})$<br>$= \alpha \cdot \beta + C_{in}(\alpha + \beta) = \boldsymbol{C_{out}}$ |

Table 17: The implementation algorithm of the proposed $PPU_6$ in the signed array multiplier.

| Step | Operation | Equivalent Logic |
|---|---|---|
| **1** | $S_1 = 0$ | $FALSE(S_1)$ |
| **2** | $S_2 = 0$ | $FALSE(S_2)$ |
| **3** | $S_3 = 0$ | $FALSE(S_3)$ |
| **4** | $(b \rightarrow 0) \equiv (b \rightarrow S_1) = S_1'$ | $S_1' = \bar{b}$ |
| **5** | $(a \rightarrow S_1') = S_1''$ | $S_1'' = \bar{a} + \bar{b} = \alpha$ |
| **6** | $(\beta \rightarrow 0) \equiv (\beta \rightarrow S_2) = S_2'$ | $S_2' = \bar{\beta}$ |
| **7** | $(S_1'' \rightarrow S_2') = S_2''$ | $S_2'' = ab + \bar{\beta}$ |
| **8** | $(S_1'' \rightarrow 0) \equiv (S_1'' \rightarrow S_3) = S_3'$ | $S_3' = ab = \bar{\alpha}$ |
| **9** | $S_1 = 0$ | $FALSE(S_1)$ |
| **10** | $(S_2'' \rightarrow 0) \equiv (S_2'' \rightarrow S_1) = S_1'$ | $S_1' = (\bar{a} + \bar{b}) \cdot \beta$ |
| **11** | $(S_3' \rightarrow \beta) = \beta'$ | $\beta' = (\bar{a} + \bar{b}) + \beta$ |
| **12** | $(\beta' \rightarrow S_1') = S_1''$ | $S_1'' = ab \cdot \bar{\beta} + (\bar{a} + \bar{b}) \cdot \beta$ |
| **13** | $S_3 = 0$ | $FALSE(S_3)$ |
| **14** | $(C_{in} \rightarrow 0) \equiv (C_{in} \rightarrow S_3) = S_3'$ | $S_3' = \overline{C_{in}}$ |
| **15** | $(S_1'' \rightarrow S_3') = S_3''$ | $S_3'' = \left((\bar{a} + \bar{b}) + \beta\right) \cdot (ab + \bar{\beta}) + \overline{C_{in}}$ |
| **16** | $S_1 = 0$ | $FALSE(S_1)$ |
| **17** | $(S_3'' \rightarrow 0) \equiv (S_3'' \rightarrow S_1) = S_1'$ | $S_1' = \left(ab \cdot \bar{\beta} + (\bar{a} + \bar{b}) \cdot \beta\right) \cdot C_{in}$ <br> $= (\bar{\alpha} \cdot \bar{\beta} + \alpha \cdot \beta) \cdot C_{in}$ |
| **18** | $(S_2'' \rightarrow C_{in}) = C_{in}'$ | $C_{in}' = (\bar{a} + \bar{b}) \cdot \beta + C_{in}$ |
| **19** | $S_2 = 0$ | $FALSE(S_2)$ |
| **20** | $(C_{in}' \rightarrow 0) \equiv (C_{in}' \rightarrow S_2) = S_2'$ | $S_2' = (ab + \bar{\beta}) \cdot \overline{C_{in}}$ |
| **21** | $(\beta' \rightarrow S_2') = S_2''$ | $S_2'' = ab \cdot \bar{\beta} + (ab + \bar{\beta}) \cdot \overline{C_{in}}$ <br> $= \bar{\alpha} \cdot \bar{\beta} + (\bar{\alpha} + \bar{\beta}) \cdot \overline{C_{in}}$ |
| **22** | $(\beta' \rightarrow C_{in}') = C_{in}''$ | $C_{in}'' = ab \cdot \bar{\beta} + \left((\bar{a} + \bar{b}) \cdot \beta + C_{in}\right)$ <br> $= \bar{\alpha} \cdot \bar{\beta} + (\alpha \cdot \beta + C_{in})$ |
| **23** | $(C_{in}'' \rightarrow S_1') = S_1''$ | $S_1'' = \left[(\alpha + \beta) \cdot \left((\bar{\alpha} + \bar{\beta}) \cdot \overline{C_{in}}\right)\right] + (\overline{\alpha \oplus \beta} \cdot C_{in})$ <br> $= (\alpha \oplus \beta \cdot \overline{C_{in}}) + (\overline{\alpha \oplus \beta} \cdot C_{in})$ <br> $= \alpha \oplus \beta \oplus C_{in} = \boldsymbol{Sum}$ |
| **24** | $S_3 = 0$ | $FALSE(S_3)$ |
| **25** | $(S_2'' \rightarrow 0) \equiv (S_2'' \rightarrow S_3) = S_3'$ | $S_3' = (\alpha + \beta) \cdot (\alpha \cdot \beta + C_{in})$ <br> $= \alpha \cdot \beta + C_{in}(\alpha + \beta) = \boldsymbol{C_{out}}$ |

As mentioned earlier, the NAND gate output is the input(s) of the full adder cell in $PPU_5$ and $PPU_6$. Therefore, it is impossible to overlap the computational steps of the NAND gate and the full adder cell in the proposed serial n-bit signed array multiplier. The output equations of $PPU_5$ and $PPU_6$ are written in (17)–(20), respectively. The IMPLY-based $PPU_5$ and $PPU_6$ serial implementation algorithms are tabulated in Tables 16 and 17, respectively.

$$\textbf{PPU}_\mathbf{5}\text{: } Sum_{FA} = \overline{ab} \oplus \overline{cd} \oplus C_{in} \equiv \left[\left(\overline{(a \to \bar{b})} \to (c \to \bar{d})\right) \to \left(\left((a \to \bar{b}) \to \overline{(c \to \bar{d})}\right) \to C_{in}\right)\right]$$

$$\to \left[\overline{\left(\left(\overline{(a \to \bar{b})} \to (c \to \bar{d})\right) \to \overline{\left((a \to \bar{b}) \to \overline{(c \to \bar{d})}\right)}\right) \to \overline{C_{in}}}\right] \tag{17}$$

$$Cout_{FA} = \overline{ab} \cdot \overline{cd} + C_{in}\left(\overline{ab} + \overline{cd}\right)$$

$$\equiv \overline{\left[\left(\overline{(a \to \bar{b})} \to (c \to \bar{d})\right) \to \overline{\left(\left((a \to \bar{b}) \to \overline{(c \to \bar{d})}\right) \to C_{in}\right)}\right]} \tag{18}$$

$$\textbf{PPU}_\mathbf{6}\text{: } Sum_{FA} = \overline{ab} \oplus \beta \oplus C_{in} \equiv \left[\left(\overline{(a \to \bar{b})} \to \beta\right) \to \left(\left((a \to \bar{b}) \to \bar{\beta}\right) \to C_{in}\right)\right]$$

$$\to \left[\overline{\left(\left(\overline{(a \to \bar{b})} \to \beta\right) \to \overline{\left((a \to \bar{b}) \to \bar{\beta}\right)}\right) \to \overline{C_{in}}}\right] \tag{19}$$

$$Cout_{FA} = \overline{ab} \cdot \beta + C_{in}\left(\overline{ab} + \beta\right) \equiv \overline{\left[\left(\overline{(a \to \bar{b})} \to \beta\right) \to \left(\left(\left((a \to \bar{b}) \to \bar{\beta}\right) \to C_{in}\right) \to 0\right)\right]} \tag{20}$$

Using fixed inputs (constant bit '1') based on the Baugh-Wooley algorithm has made it possible to modify the output logic functions of the half adder ($PPU_3$) and the full adder ($PPU_8$). As a result, the number of computational steps of these two PPUs based on the IMPLY logic in the serial architecture is reduced to its minimum. If one of the inputs in (2) and (3) is considered equal to '1', the output functions are changed to (21) and (22).

$$Sum_{HA} = \alpha \oplus \beta = 1 \oplus \beta = \bar{\beta} \tag{21}$$

$$Cout_{HA} = \beta \cdot \alpha = \beta \tag{22}$$

In the classic structure, the implementation algorithm of the half adder cell takes twelve computational steps, but in the proposed structure of $PPU_3$, the number of computational steps is reduced to two steps (83% improvement). The implementation algorithm of the proposed $PPU_3$ is given in Table 18. According to the output equations of the full adder (see (6) and (7)), if one of the inputs of this cell is a constant bit '1', then the computational complexity of the output equations is reduced in this cell. By applying the constant bit '1' to (6) and (7), these output equations convert to (23) and (24), which are $PPU_8$'s outputs. These outputs can be implemented serially based on the IMPLY logic in nine computational steps (59% improvement) with four memristors (two input memristors and two work memristors), as presented in Table 19.

$$Sum_{FA} = x \oplus y \oplus C_{in} = 1 \oplus y \oplus C_{in} = \overline{y \oplus C_{in}} \tag{23}$$

$$Cout_{FA} = x \cdot y + C_{in}(x + y) = y + C_{in} \tag{24}$$

The proposed n-bit signed CSA-based array multiplier's computational steps have been reduced from $(27n^2 - 24n + 24)$ to $(25n^2 - 32n + 1)$ based on the proposed implementation algorithms of $PPU_1$–$PPU_8$. The equation used to calculate the computational steps of the proposed signed array multiplier is written in (25). $CS_{SAmultiplier}$ refers to the number of computational steps of the signed array multiplier in (25). Furthermore, $5n - 4$ memristors are required to implement the n-bit signed CSA-based array multiplier, the same as its classic design.

$$\begin{aligned} CS_{SAmultiplier} = {} & \#PPU_1 \times CS_{PPU1} + \#PPU_2 \times CS_{PPU2} + \#PPU_3 \times CS_{PPU3} + \#PPU_4 \times CS_{PPU4} \\ & + \#PPU_5 \times CS_{PPU5} + \#PPU_6 \times CS_{PPU6} + \#PPU_7 \times CS_{PPU7} + \#PPU_8 \times CS_{PPU8} \\ & + \#FA \times CS_{FA} + \#AND \times CS_{AND} \end{aligned} \tag{25}$$

Table 18: The implementation algorithm of the proposed $PPU_3$ in the signed array multiplier.

| Step | Operation | Equivalent Logic |
|---|---|---|
| 1 | $S_1 = 0$ | $FALSE(S_1)$ |
| 2 | $(\beta \to 0) \equiv (\beta \to S_1) = S_1{}'$ | $S_1{}' = \bar{\beta} = \boldsymbol{Sum}$ |

Table 19: The implementation algorithm of the proposed $PPU_8$ in the signed array multiplier.

| Step | Operation | Equivalent Logic |
|---|---|---|
| 1 | $S_1 = 0$ | $FALSE(S_1)$ |
| 2 | $S_2 = 0$ | $FALSE(S_2)$ |
| 3 | $(C_{in} \rightarrow 0) \equiv (C_{in} \rightarrow S_1) = S_1{}'$ | $S_1{}' = \overline{C_{in}}$ |
| 4 | $(\beta \rightarrow S_1{}') = S_1{}''$ | $S_1{}'' = \bar{\beta} + \overline{C_{in}}$ |
| 5 | $(\beta \rightarrow 0) \equiv (\beta \rightarrow S_2) = S_2{}'$ | $S_2{}' = \bar{\beta}$ |
| 6 | $(S_2{}' \rightarrow C_{in}) = C_{in}{}'$ | $C_{in}{}' = \beta + C_{in} = \boldsymbol{C_{out}}$ |
| 7 | $S_2 = 0$ | $FALSE(S_2)$ |
| 8 | $(S_1{}'' \rightarrow 0) \equiv (S_1{}'' \rightarrow S_2) = S_2{}'$ | $S_2{}' = \beta \cdot C_{in}$ |
| 9 | $(C_{in}{}' \rightarrow S_2{}') = S_2{}''$ | $S_2{}'' = \bar{\beta} \cdot \overline{C_{in}} + \beta \cdot C_{in} = \overline{\beta \oplus C_{in}} = \boldsymbol{Sum}$ |

## 4 Simulation results, evaluation, and comparison

This section of the article has been presented in two subsections. The first subsection presents the simulation results of the proposed structures at the circuit level. It also includes a comprehensive comparison of the area, energy consumption, and computational delay of the proposed multipliers with previous works, along with an evaluation of the results. In the second part, the functionality of the proposed multipliers is assessed in the applications of Gaussian blur and edge detection. The results of hardware implementation and computational complexity of these applications based on the proposed multipliers are compared with SOA.

### 4.1 Circuit simulation and its results

Various SPICE models have been presented to simulate memristive circuits. Voltage ThrEshold Adaptive Memristor (VTEAM) model is one of the nonlinear memristive models with high computational accuracy and low computational complexity [1, 31, 32]. This research applied the VTEAM model introduced at *TU Wien* to simulate the proposed circuits [4–6, 20–22, 28, 30]. The specifications of the applied memristor model and the IMPLY logic's design parameters are detailed in Table 20. In the circuit-level simulation, the pulse width of each computational step is considered equal to 30 $\mu s$, as reported in [4–6, 20–22, 28, 30].

Table 20: Setup values of IMPLY logic and VTEAM model [4, 5, 22, 28, 30].

| Parameter | $v_{set}$ | $v_{reset}$ | $v_{cond}$ | $t_{pulse}$ | $R_G$ | $v_{off}$ | $v_{on}$ | $\alpha_{off}$ | $\alpha_{on}$ |
|---|---|---|---|---|---|---|---|---|---|
| **Value** | 1 V | 1 V | 900 mV | 30 $\mu s$ | 40 KΩ | 0.7 V | -10 mV | 3 | 3 |
| **Parameter** | $R_{off}$ | $R_{on}$ | $k_{off}$ | $k_{on}$ | $w_{off}$ | $w_{on}$ | $w_C$ | $a_{off}$ | $a_{on}$ |
| **Value** | 1 MΩ | 10 KΩ | 1 $\frac{cm}{s}$ | -0.5 $\frac{nm}{s}$ | 0 nm | 3 nm | 107 pm | 3 nm | 0 nm |

The proposed PPUs introduced in the last section and other PPUs of unsigned and signed array multipliers are simulated in the SPICE simulator using the mentioned VTEAM model. All the possible input states of the proposed circuits are simulated to ensure their correct functionality. For example, an unsigned array multiplier's $PPU_1$ has four inputs, so 16 input states must be assessed. All the outputs of PPUs were calculated correctly in all cases based on the simulation results. Only the simulation waveforms of one input state of the proposed unsigned array multiplier's $PPU_{1\text{-}3}$ are presented in Figures 4–6 due to the multiplicity of PPUs and their input states.

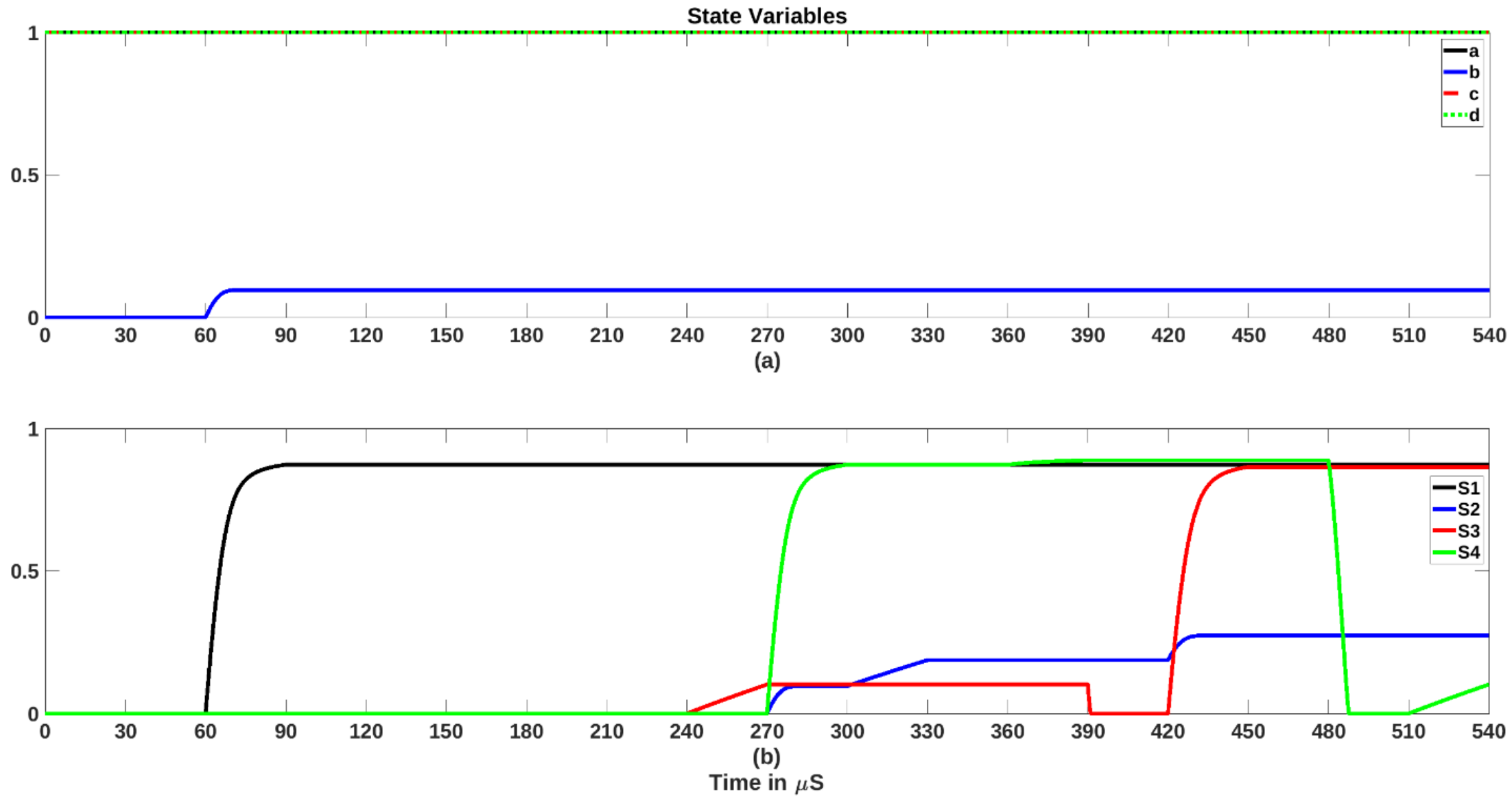

Figure 4: Simulation waveforms of the proposed unsigned array multiplier's *PPU₁* for input state "1011".

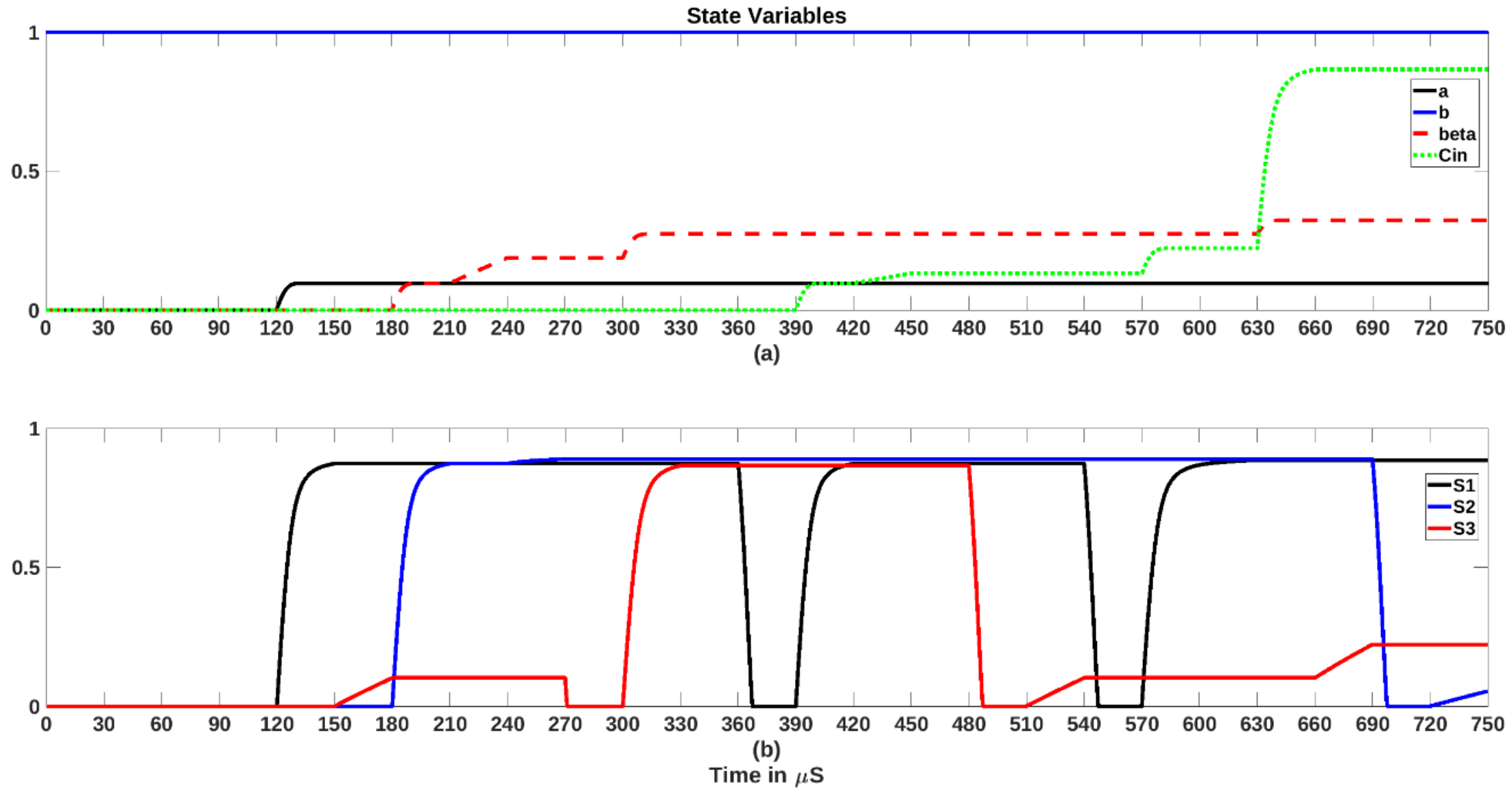

Figure 5: Simulation waveforms of the proposed unsigned array multiplier's $PPU_2$ for input state "0001".

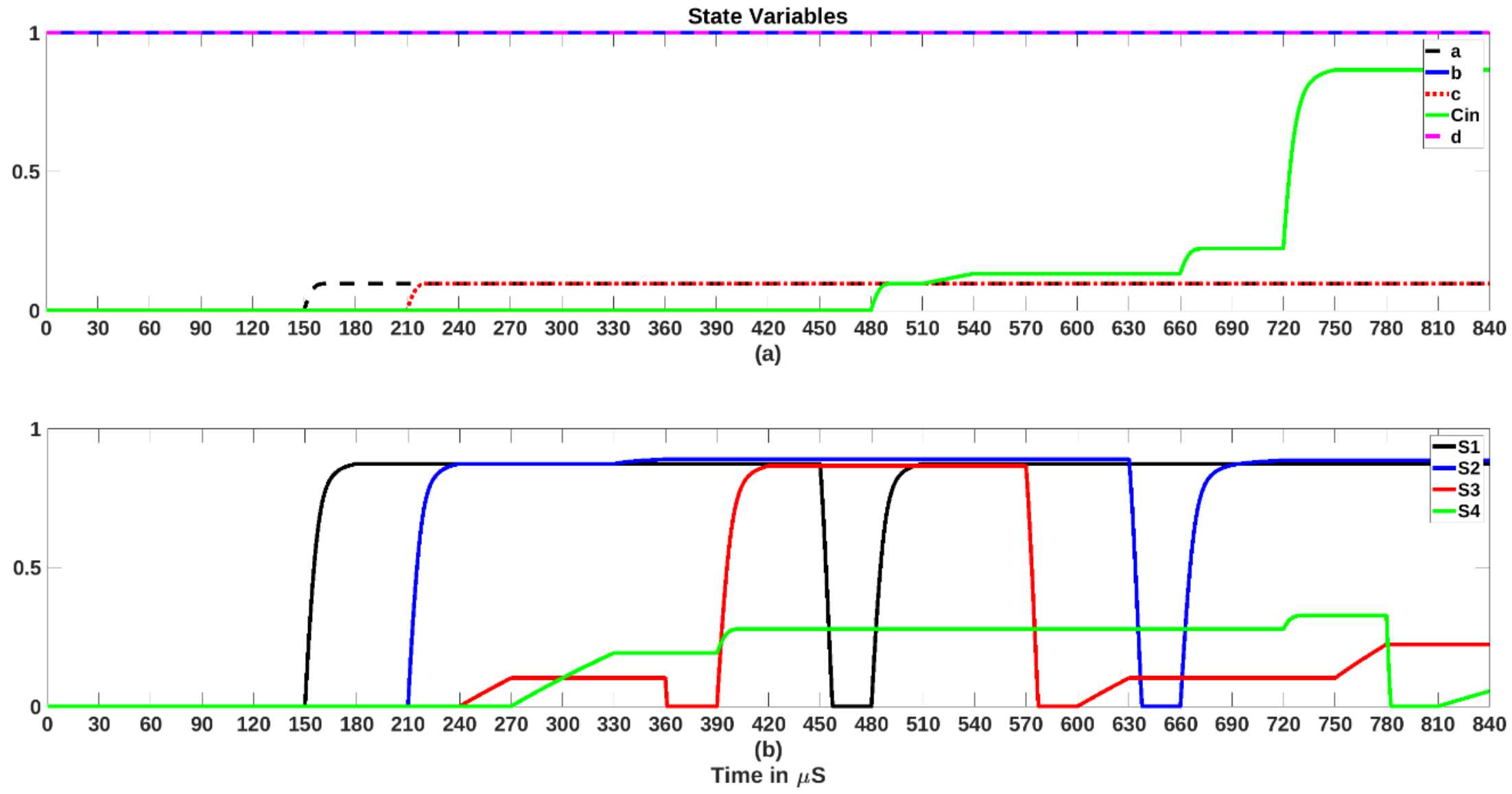

Figure 6: Simulation waveforms of the proposed unsigned array multiplier's $PPU_3$ for input state "00101".

The output waveform of the proposed signed array multiplier's PPUs ($PPU_2$ and $PPU_7$) for one of their input states are illustrated in Figures 7 and 8.

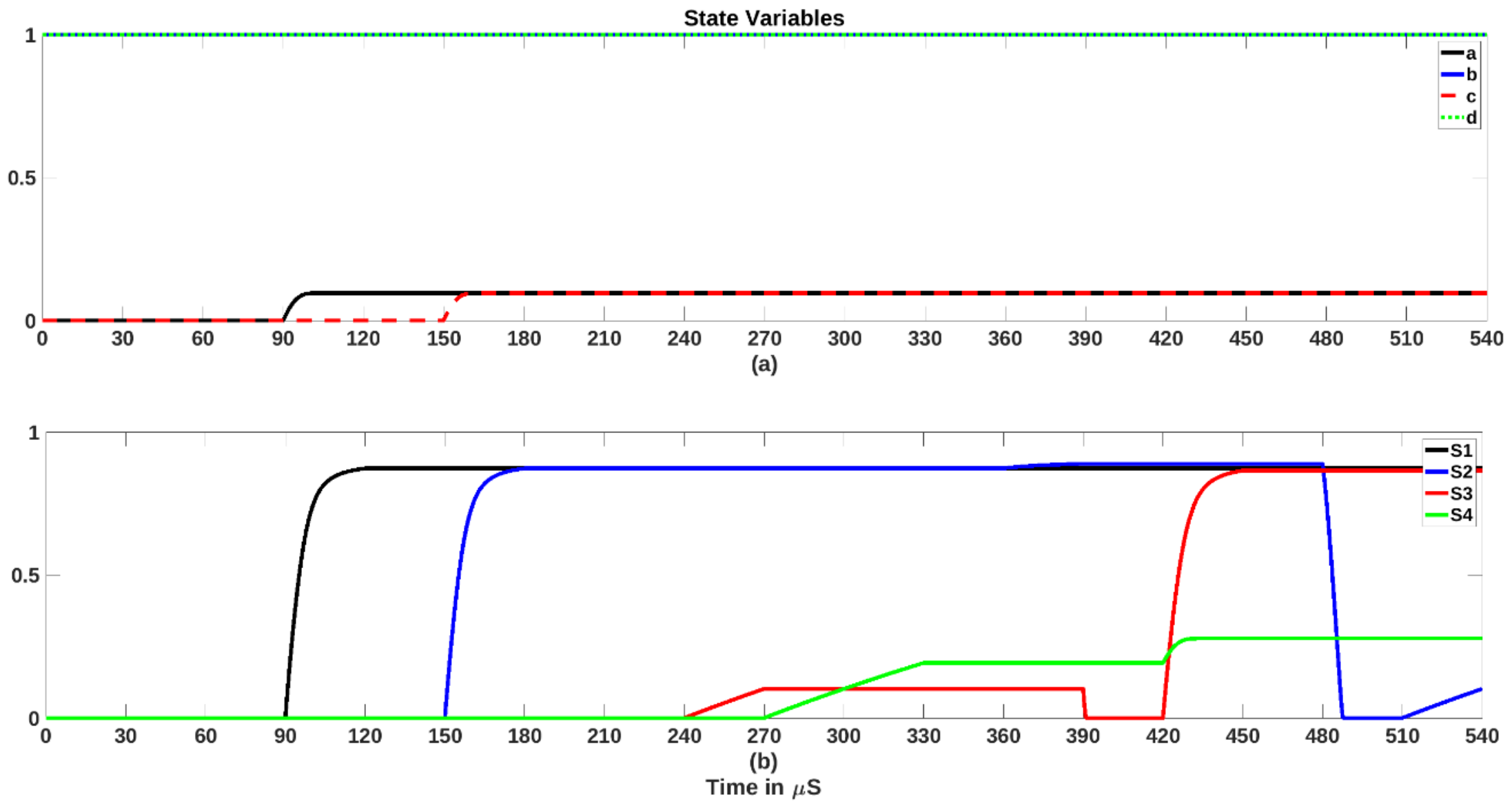

Figure 7: Simulation waveforms of the proposed signed array multiplier's $PPU_2$ for input state "0101".

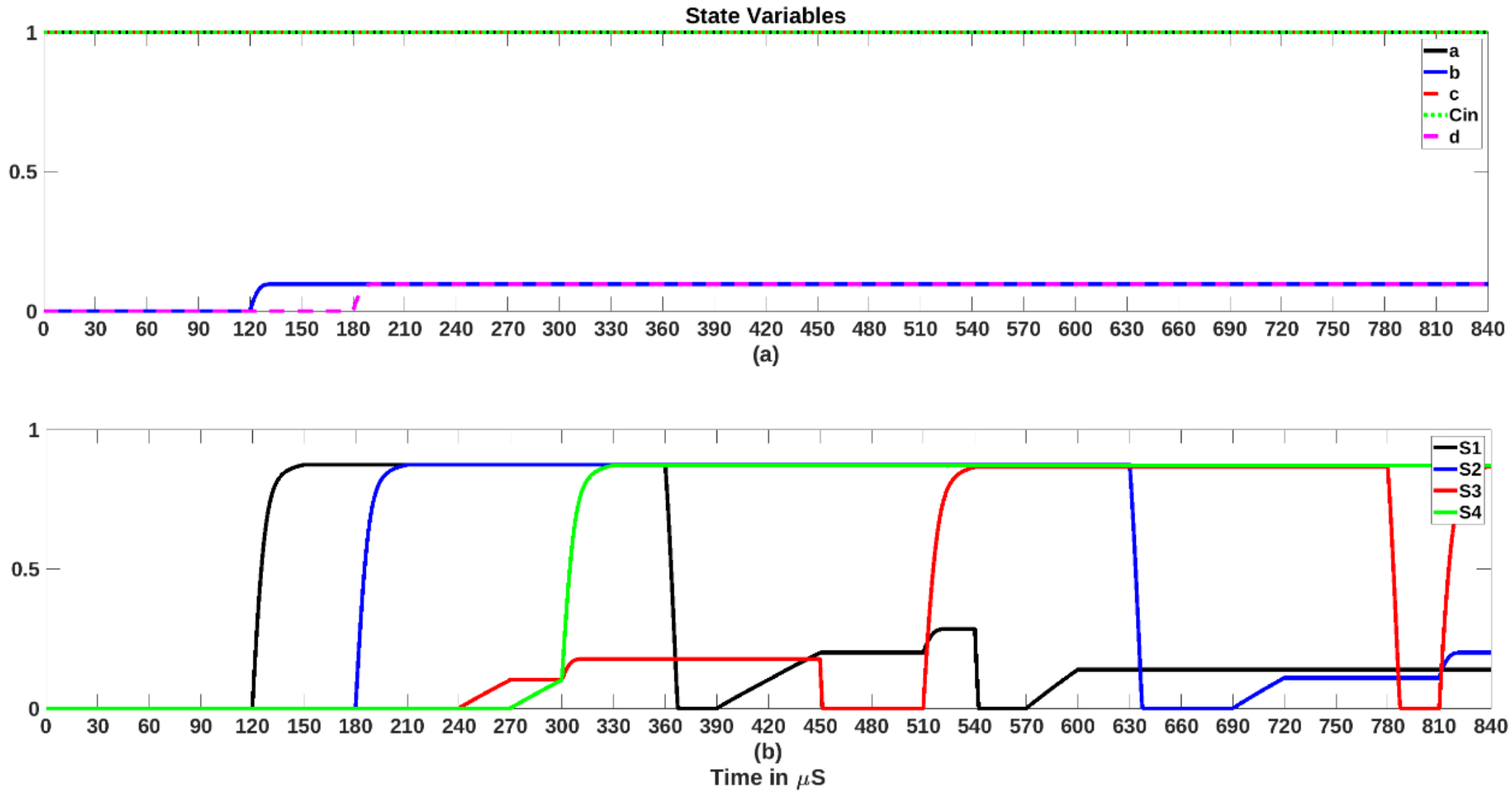

Figure 8: Simulation waveforms of the proposed signed array multiplier's $PPU_7$ for input state "11010".

Proposed $PPU_{1\text{-}3}$ of the unsigned array multiplier were implemented in 18, 25, and 28 computational steps, simulated in 0 µs–540 µs, 0 µs–750 µs, and 0 µs–850 µs, respectively. 450 µs–480 µs, 660 µs–690 µs, and 750 µs–780 µs are the intervals to compute the *Sum* output of the proposed $PPU_{1\text{-}3}$, respectively, in the unsigned array multiplier structure. The $C_{out}$ output is computed and stored at the last computational step in all three proposed PPUs of this multiplier. Proposed $PPU_2$ and $PPU_7$ of the signed array multiplier are simulated in the time intervals of 0 µs–540 µs (18 computational steps) and 0 µs–840 µs (28 computational steps), respectively. The *Sum* output of the proposed $PPU_2$ and $PPU_7$ is simulated in the intervals of 450 µs–480 µs and 750 µs–780 µs, respectively, and each PPU's $C_{out}$ output is calculated and stored in the last computational step.

The method introduced in [4, 5, 28, 30] was applied to calculate the energy consumption of each cell of the proposed unsigned and signed array multipliers. In Table 21, the estimated energy consumption of each component of the proposed unsigned and signed array multipliers is written along with the constituent units of other multipliers (such as the multiplexer in the radix-2 Booth multiplier and the 4:2 compressor proposed in [28]). The energy consumption of unsigned and signed array multipliers has been calculated by applying (26) and (27), respectively. $E_{UAmultiplier}$ and $E_{SAmultiplier}$ refer to the energy consumption of the unsigned and signed array multipliers in these equations. According to (26) and (27), the estimated energy consumption of n-bit unsigned and signed array multipliers equal $(2.156n^2 - 2.672n - 0.022)$ nJ and $(2.156n^2 - 2.703n - 0.067)$ nJ, respectively.

Table 21: The estimated energy consumption of the operations and computing elements applied in the proposed unsigned and signed multipliers [4, 5, 22, 28].

| Operation/ Component | Energy Consumption (nJ) | Operation/ Component | Energy Consumption (nJ) | Operation/ Component | Energy Consumption (nJ) |
|---|---|---|---|---|---|
| FALSE | 0.05 | NOT/Proposed Signed $PPU_3$ | 0.13 | AND | 0.33 |
| NAND | 0.24 | OR | 0.244 | XOR | 0.374 |
| XNOR | 0.9 | First MUX 2:1 | 0.6 | Second MUX 2:1 | 0.9 |
| Half Adder [28] | 1.02 | Full Adder [23] | 1.85 | 4:2 Compressor [28] | 3.76 |
| Classic Unsigned $PPU_1$ | 1.68 | Classic Unsigned $PPU_2$ | 2.18 | Classic Unsigned $PPU_3$ | 2.51 |
| Classic Signed $PPU_1$ | 1.68 | Classic Signed $PPU_2$ | 1.59 | Classic Signed $PPU_3$ | 1.02 |
| Classic Signed $PPU_4$ | 2.18 | Classic Signed $PPU_5$ | 2.33 | Classic Signed $PPU_6$ | 2.09 |
| Classic Signed $PPU_7$ | 2.42 | Classic Signed $PPU_8$ | 1.85 | Proposed Unsigned $PPU_1$ | 1.602 |
| Proposed Unsigned $PPU_2$ | 2.156 | Proposed Unsigned $PPU_3$ | 2.5 | Proposed Signed $PPU_1$ | 1.602 |
| Proposed Signed $PPU_2$ | 1.62 | Proposed Signed $PPU_4$ | 2.156 | Proposed Signed $PPU_5$ | 2.5 |
| Proposed Signed $PPU_6$ | 2.15 | Proposed Signed $PPU_7$ | 2.475 | Proposed Signed $PPU_8$ | 0.74 |

$$E_{UAmultiplier} = \#PPU_1 \times E_{PPU1} + \#PPU_2 \times E_{PPU2} + \#PPU_3 \times E_{PPU3} + \#HA \times E_{HA} + \#FA \times E_{FA} + \#AND \times E_{AND} \tag{26}$$

$$E_{SAmultiplier} = \#PPU_1 \times E_{PPU1} + \#PPU_2 \times E_{PPU2} + \#PPU_3 \times E_{PPU3} + \#PPU_4 \times E_{PPU4} + \#PPU_5 \times E_{PPU5} + \#PPU_6 \times E_{PPU6} + \#PPU_7 \times E_{PPU7} + \#PPU_8 \times E_{PPU8} + \#FA \times E_{FA} + \#AND \times E_{AND} \tag{27}$$

Table 22 compares the proposed serial unsigned array multiplier with other classic serial unsigned multipliers regarding the number of memristors, computational steps, and energy consumption. Furthermore, the proposed serial signed array multiplier is evaluated and compared with the classic signed multipliers, and the results are reported in Table 23. The same method (see (12) and (25)–(27)) is applied to compute the number of computational steps and energy consumption of classic multipliers. The authors tried to reduce the number of applied work memristors to its minimum in the serial unsigned and signed array multipliers to calculate the number of required memristors.

Table 22: The comparison of IMPLY-based serial unsigned multipliers.

| Multiplier | No. of Steps | | | No. of Memristors | | | Energy Consumption (nJ) | | |
|---|---|---|---|---|---|---|---|---|---|
| | Total | n=8 | %Imp. | Total | n=8 | %Imp. | Total | n=8 | %Imp. |
| Dadda | $27n^2 - 32n$ | 1472 | 26.25 | $n^2 + 2$ | 66 | - | $2.18n^2 - 2.68n$ | 118.08 | 29.7 |
| [28] | $27n^2 - 32n$ | 1472 | 26.25 | $n^2 + 2$ | 66 | - | $2.21n^2 - 2.8n - 0.05$ | 119 | 29.15 |
| Add & Shift | $31n^2 + n + 4$ | 1996 | - | $3n + 5$ | 29 | 56.07 | $2.623n^2 - 0.023n + 0.26$ | 167.95 | - |
| Array Multiplier (Classic) | $27n^2 - 32n$ | 1472 | 26.25 | $5n - 4$ | 36 | 45.46 | $2.18n^2 - 2.68n$ | 118.08 | 29.7 |
| Array Multiplier (Proposed) | $25n^2 - 32n + 2$ | 1346 | 32.57 | $5n - 4$ | 36 | 45.46 | $2.156n^2 - 2.672n - 0.022$ | 116.59 | 30.58 |

According to the results of Table 22, the proposed unsigned and signed array multipliers compute the final result in fewer cycles than SOA. The number of computational steps of the proposed 4-bit unsigned array multiplier is reduced by 46% compared to the classic 4-bit add & shift multiplier and by 10% compared to the other three classic serial multipliers. In the 8-bit architecture, the proposed unsigned array multiplier reduces the number of computational steps by at least 9% (compared to classic unsigned array and Dadda multipliers and the proposed multiplier in [28]) and a maximum of 36% (compared to the unsigned add & shift multiplier). The most energy-efficient architecture among the unsigned serial multipliers investigated in this article is the proposed array multiplier. Due to the reduction of computational steps and memristors involved in implementing the multiplier's structure, an improvement of 1.5%–44% in energy consumption has been achieved compared to classic designs (4-bit and 8-bit architectures). The proposed 4-bit serial unsigned multiplier requires the least number of memristors to compute the final result as it is improved by 6%–11% compared to classic unsigned multipliers. The classic 8-bit unsigned add & shift multiplier requires only 29 memristors for its implementation, which is the least number of memristors compared to other classic unsigned multipliers and the proposed one. The proposed multiplier has improved the number of memristors required for implementing in the crossbar array structure by 47% compared to the classic unsigned Dadda multiplier and the proposed multiplier in [28].

Table 23 compares the proposed signed array multiplier with signed add & shift, radix-2 Booth, Dadda (based on the Baugh-Wooley method), and classic signed array multipliers. The proposed signed array multiplier takes the least computational steps to calculate the output results compared to other memristive serial signed multipliers. Accordingly, compared to the classic signed architectures, the proposed 4-bit and 8-bit multipliers have improved the computational steps by 6%–65% and 7%–59%, respectively. Energy consumption is another essential circuit evaluation criterion that gets the attention of designers and researchers. The proposed signed array multiplier consumes the lowest energy compared to other classic memristive signed multipliers, and the radix-2 Booth multiplier consumes the highest energy. By reducing the number of computational steps and the involved work memristors, dynamic energy consumption is improved. Thus, the proposed 8-bit multiplier reduces the energy consumption by a maximum of 57% (compared to the radix-2 Booth multiplier) and a minimum of 2% (compared to the classic signed array multiplier). The proposed architecture requires only 16 memristors to implement the 4-bit signed array multiplier, improved by 6%–33% compared to other serial multipliers in this structure. The number of required memristors of the proposed 8-bit array multiplier is 19% more than the signed add & shift multiplier. However, it is 10% and 45% less than the radix-2 Booth multiplier and the classic Dadda multiplier. The add & shift multiplier has a simple architecture, so the number of memristors in this multiplier is minimal. However, the number of computational steps and energy consumption in this multiplier is high.

Table 23: The comparison of IMPLY-based serial signed multipliers.

| Multiplier | No. of Steps | | | No. of Memristors | | | Energy Consumption (nJ) | | |
|---|---|---|---|---|---|---|---|---|---|
| | Total | n=8 | %Imp. | Total | n=8 | %Imp. | Total | n=8 | %Imp. |
| Add & Shift | $31n^2 + 6n + 9$ | 2041 | 26.25 | $3n + 5$ | 29 | 56.06 | $2.623n^2 + 0.287n + 0.57$ | 170.74 | 37.47 |
| Radix-2 Booth | $49n^2 + 15n - 4$ | 3252 | 26.25 | $4n + 8$ | 40 | 39.4 | $4.169n^2 + 0.804n - 0.2$ | 273.05 | - |
| Baugh-Wooley (Classic) | $27n^2 - 24n + 24$ | 1560 | - | $n^2 + 2$ | 66 | - | $2.18n^2 + 1.84n + 1.63$ | 126.43 | 53.7 |
| Array Multiplier (Classic) | $27n^2 - 36n + 3$ | 1443 | 26.25 | $5n - 4$ | 36 | 45.46 | $2.18n^2 - 2.86n + 2.03$ | 118.67 | 56.54 |
| Array Multiplier (Proposed) | $25n^2 - 32n + 1$ | 1345 | 32.57 | $5n - 4$ | 36 | 45.46 | $2.156n^2 - 2.073n - 0.067$ | 116.29 | 57.41 |

## 4.2 Application-level simulation and its results

Image processing is one of the data-intensive applications that has received much attention these days and plays a significant role in daily human lives. Slipping special kernels (filters) on reference images and performing convolution operations between the pixels of the image and the kernel are applied in various applications [33]. Applying the Gaussian blur kernel is one of the methods that can be applied to reduce the noise of the reference image (see (28)).

$$\omega = \frac{1}{16}\begin{bmatrix} 1 & 2 & 1 \\ 2 & 4 & 2 \\ 1 & 2 & 1 \end{bmatrix} \quad (28)$$

Table 24: The number of applied operations to the 3*3 Gaussian blur kernel on an n*n pixels grayscale image.

| **Arithmetic Operation** | **No. of Operations** | **n=256** |
|---|---|---|
| 8-bit unsigned multiplier | $9n^2 - 36n + 36$ | 580644 |
| 8-bit unsigned adder | $2n^2 - 8n + 8$ | 129032 |
| 9-bit unsigned adder | $3n^2 - 12n + 12$ | 193548 |
| 10-bit unsigned adder | $2n^2 - 8n + 8$ | 129032 |
| 11-bit unsigned adder | $n^2 - 4n + 4$ | 64516 |
| Divider | 1 | 1 |

Table 25: The number of applied operations to the 3*3 edge detection kernel on an n*n pixels grayscale image.

| **Arithmetic Operation** | **No. of Operations** | **n=256** |
|---|---|---|
| 8-bit signed multiplier | $5n^2 - 20n + 20$ | 322580 |
| 9-bit signed adder | $2n^2 - 8n + 8$ | 129032 |
| 10-bit signed adder | $n^2 - 4n + 4$ | 64516 |
| 11-bit unsigned adder | $n^2 - 4n + 4$ | 64516 |

In a 3*3 Gaussian blur filter, nine unsigned multiplication operations along with several unsigned addition operations (using 8-, 9-, 10-, and 11-bit adders) should be done to apply this kernel to the image. Also, a division (shift) operation is required for this application. Table 24 reports the number of arithmetic operations required to apply the 3*3 Gaussian blur kernel on an n*n pixels grayscale image (8-bit).
Another proper kernel in image processing is written in (29). This kernel is applied to detect the edges of an image. Edge detection is one of the essential operations in the application of pattern recognition [32]. Signed multipliers are needed to detect edges based on the kernel in (29).

$$\omega = \begin{bmatrix} 0 & -1 & 0 \\ -1 & 4 & -1 \\ 0 & -1 & 0 \end{bmatrix} \tag{29}$$

The edge detection kernel slides over the reference image, and the convolution result is stored in the corresponding pixel. This convolution operation needs five signed multipliers and some 9-, 10- and 11-bit adders. Table 25 specifies the number of multiplication and addition operations required to detect the edges of an n*n grayscale image.
The behavioral simulation results of Gaussian blur and edge detection applications are depicted in Figure 9. The number of computational steps, memristors, and energy consumption of the unsigned and signed multipliers applied for Gaussian blur and edge detection kernels are presented in Table 26.

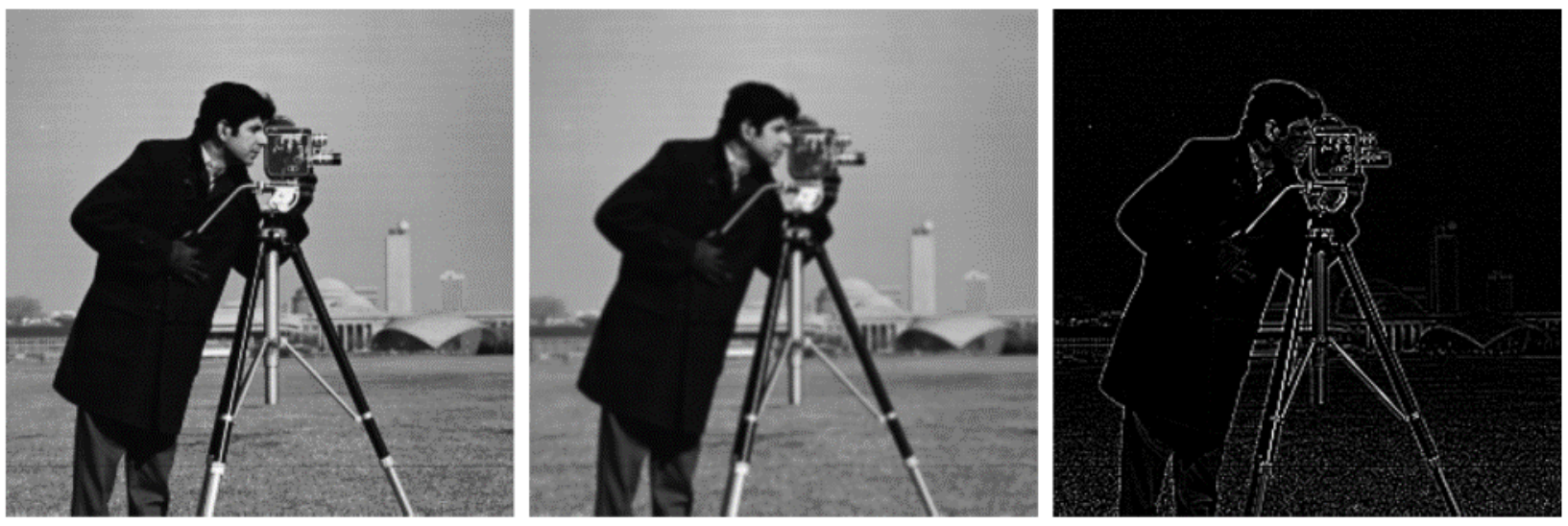

**(a)** **(b)** **(c)**

Figure 9: Simulation result of: b) Gaussian blur, and c) edge detection.

The authors' main contribution is to propose memristive unsigned and signed array multipliers. Hence, only the results of the multiplication operations in Gaussian blur and edge detection applications on the reference image (see Figure 9. a) have been reported and compared. It should be noted that the number of computational steps, memristors, and energy consumption of other basic arithmetic operations (e.g., addition) are considered the same for these two image-processing applications in all architectures. According to the results of Table 26, the number of computational steps, and energy consumption 26%–33%, and 29%–31% are improved, respectively, by applying the proposed unsigned array multiplier in the Gaussian blur application (256*256 grayscale image). In the application of edge detection (256*256 grayscale image), the number of computational steps, and energy consumption are reduced by at least 37%, and 37%, respectively, based on the proposed signed array multiplier.

Table 26: Evaluation of the unsigned and signed multipliers applied for Gaussian blur and edge detection kernels.

| Multiplier | No. of Computational steps | No. of Memristors | Energy Consumption ($\times 10^{-3} J$) |
|---|---|---|---|
| **Gaussian Blur** | | | |
| Dadda | $580644 \times 1472 = 854707968$ | $580644 \times 16 + 50 = 9290354$ | $580644 \times 118.08nJ = 68.562$ |
| [28] | $580644 \times 1472 = 854707968$ | $58064416 + 50 = 9290354$ | $580644 \times 119nJ = 69.096$ |
| Add & Shift | $580644 \times 1996 = 1158965424$ | $580644 \times 16 + 12 = 9290316$ | $580644 \times 167.948nJ = 97.517$ |
| Classic Array | $580644 \times 1472 = 854707968$ | $580644 \times 16 + 20 = 9290324$ | $580644 \times 118.08 = 68.562$ |
| Proposed | $580644 \times 1346 = 781546824$ | $580644 \times 16 + 20 = 9290324$ | $580644 \times 116.586nJ = 67.694$ |
| **Edge Detection** | | | |
| Add & Shift | $322580 \times 2574 = 830320920$ | $322580 \times 18 + 14 = 5806454$ | $322580 \times 216.389nJ = 69.802$ |
| Booth (Radix-2) | $322580 \times 4100 = 1322578000$ | $322580 \times 18 + 26 = 5806466$ | $322580 \times 344.725nJ = 111.201$ |
| Baugh-Wooley (Classic) | $322580 \times 1995 = 643547100$ | $322580 \times 18 + 65 = 5806505$ | $322580 \times 161.65nJ = 52.145$ |
| Classic Array | $322580 \times 1866 = 601934280$ | $322580 \times 18 + 23 = 5806463$ | $322580 \times 151.76nJ = 48.954$ |
| Proposed | $322580 \times 1738 = 560644040$ | $322580 \times 18 + 23 = 5806463$ | $322580 \times 150.242nJ = 48.465$ |

# 5 Conclusion

The design of efficient memristive arithmetic units, such as multipliers, for processing data-intensive applications in memory is of great importance today. In this research, the authors have focused on the design of multipliers based on the IMPLY design method for PIM. Overlapping the computational steps and presenting new crossbar array-friendly IMPLY-based implementation algorithms for PPUs is the major contribution of the authors. Moreover, the unsigned and signed array multipliers are redesigned by applying the proposed PPUs. The reduction of computational steps and energy consumption of the proposed PPUs compared to similar classic structures has improved the execution time of the proposed multipliers and their energy consumption. The proposed 8-bit unsigned array multiplier has improved the number of computational steps by 9%–36% and energy consumption by 1.5%–31%. The number of memristors required for implementing the proposed multiplier in the crossbar array is also improved by up to 47% compared to other classic unsigned multipliers. Furthermore, the proposed 8-bit signed array multiplier has improved the number of computational steps (up to 59%), energy consumption (up to 57%), and the number of required memristors (up to 45%). Examining the functionality of the proposed multipliers in the applications of Gaussian blur and edge detection has indicated the improvement of the circuit evaluation criteria compared to the classic multipliers.

# Author Contributions

Seyed Erfan Fatemieh: Conceptualization, Methodology, Software, Validation, Formal Analysis, Investigation, Data Curation, Writing - Original Draft, Writing – Review & Editing, Visualization.
Bahareh Bagheralmoosavi: Conceptualization, Methodology, Software, Validation, Formal Analysis, Investigation, Data Curation, Writing - Original Draft, Writing - Review & Editing, Visualization.
Mohammad Reza Reshadinezhad: Software, Validation, Formal Analysis, Investigation, Resources, Writing - Review & Editing, Supervision, Project Administration.

## Acknowledgements

This research did not receive any specific grant from funding agencies in the public, commercial, or not-for-profit sectors.

## Data Availability Statement

Data is contained within the article.

## Conflicts of Interest

The authors declare no conflicts of interest.